\documentclass[useAMS,usenatbib]{mn2e}
\usepackage{graphicx}
\usepackage{amsmath}
\usepackage{amssymb}
\usepackage{color}

\title[Density structures in supersonic turbulent flows]{Mach number study of supersonic turbulence:\\The properties of the density field}
\author[L. Konstandin, W. Schmidt, P. Girichidis, T. Peters, R. Shetty, and R. S. Klessen]
{L. Konstandin\thanks{E-mail: email@LukasKonstandin.de}$^{1,\,2,}$, W. Schmidt$^3$, P. Girichidis,$^4$ T. Peters$^4$, R. Shetty$^2$, and R. S. Klessen$^{2,5,6}$\\
$^{1}$School of Physics \& Astronomy, University of Exeter, Stocker Road, Exeter EX4 4QL, UK\\
$^{2}$Universit\"{a}t Heidelberg, Zentrum f\"{u}r Astronomie, Institut f\"{u}r Theoretische Astrophysik, Albert-Ueberle-Str. 2, 69120 Heidelberg, Germany\\
$^{3}$Hamburger Sternwarte, Universit\"at Hamburg, Gojenbergsweg 112, 21029 Hamburg\\ 
$^{4}$Max-Planck-Institut f\"ur Astrophysik,  Karl-Schwarzschild-Str.~1, 85741 Garching, Germany\\
$^{5}$Department of Astronomy and Astrophysics, University of California, 1156 High Street, Santa Cruz,\\
 CA 95064, USA\\ 
$^{6}$Kavli Institute for Particle Astrophysics and Cosmology, Stanford University, SLAC National Accelerator Laboratory, Menlo Park,\\ CA 94025, USA}

\begin{document}
\date{Accepted 2015 XXX YY. Received 2015 June 10; in original form 2015 XXX YY}
\pagerange{\pageref{firstpage}--\pageref{lastpage}} \pubyear{2015}
\maketitle
\label{firstpage}

\newcommand{\AAA}{A\&A}
\newcommand{\apj}{ApJ}
\newcommand{\apjl}{ApJl}
\newcommand{\apjs}{ApJs}

\newcommand{\mnras}{MNRAS}

\newcommand{\araa}{ARAA}

\newcommand{\physrep}{Phys.~Rep.}

\newcommand{\pree}{Phy.~Rev.~E}
\newcommand{\pre}{Phy.~Rev.~E}

\newcommand{\aap}{AAP}

\newcommand{\jfm}{JFM}

\newcommand{\com}[1]{\textcolor{green}{\textbf{(#1)}}}
\newcommand{\todocom}[1]{\textcolor{red}{\textbf{(TODO ...)}}\textcolor{green}{\textbf{(#1)}}}
\newcommand{\todo}{\textcolor{red}{\textbf{(TODO ...)}}}
\newcommand{\mynew}[1]{\textcolor{red}{#1}}
%%%%%%%%%%%%%%%
%%% ABSTRACT  %
%%%%%%%%%%%%%%%
%% Not longer than 250 words
%% Not longer than 250 words
%% Not longer than 250 words
\begin{abstract}
%We analyse the scaling properties of turbulent flows using a suite of three-dimensional numerical simulations.
We model driven, compressible, isothermal, turbulence with Mach numbers ranging from the subsonic ($\mathcal{M} \approx 0.65$) to the highly supersonic regime ($\mathcal{M}\approx 16 $). The forcing scheme consists both solenoidal (transverse) and compressive (longitudinal) modes in equal parts.
We find a relation $\sigma_{s}^2 = \mathrm{b}\log{(1+\mathrm{b}^2\mathcal{M}^2)}$ between the Mach number and the standard deviation of the logarithmic density with $\mathrm{b} = 0.457 \pm 0.007$.
The density spectra follow $\mathcal{D}(k,\,\mathcal{M}) \propto k^{\zeta(\mathcal{M})}$ with scaling exponents depending on the Mach number.
We find $\zeta(\mathcal{M}) = \alpha \mathcal{M}^{\beta}$ 
with a coefficient $\alpha$ that varies slightly with resolution, whereas $\beta$ changes systematically.
We extrapolate to the limit of infinite resolution and find $\alpha = -1.91 \pm 0.01,\, \beta =-0.30\pm 0.03$.
The dependence of the scaling exponent on the Mach number implies a fractal dimension $D=2+0.96 \mathcal{M}^{-0.30}$.
We determine how the scaling parameters depend on the wavenumber and find that the density spectra are slightly curved. This curvature gets more pronounced with increasing Mach number.
We propose a physically motivated fitting formula $\mathcal{D}(k) = \mathcal{D}_0 k^{\zeta k^{\eta}}$
by using simple scaling arguments. The fit
reproduces the spectral behaviour down to scales $k\approx 80$.
The density spectrum follows a single power-law $\eta = -0.005 \pm 0.01$ in the low Mach number regime
and the strongest curvature $\eta = -0.04 \pm 0.02$ for the highest Mach number.
These values of $\eta$ represent a lower limit, as the curvature increases with resolution.
\end{abstract}

%%%%%%%%%%%%%%%%%%%%%%
%%% NOW THE KEYWORDS %
%%%%%%%%%%%%%%%%%%%%%%
%% one and six keywords
%% check list :http://oxfordjournals.org/our_journals/mnrasl/for_authors/mnraskeywords.pdf

%%%%%%%%%%%%%%%%%%%%%%%
%%% BODY OF THE PAPER %
%%%%%%%%%%%%%%%%%%%%%%%
%###############################################################################
%INTRODUCTION
%###############################################################################
\section{Introduction}
Turbulence is an extremely important phenomenon in astrophysical media on nearly all scales.
It occurs in accretion discs \citep{Meschiari2012}, supernova remnants \citep{Inoue2009}, star forming molecular clouds  \citep{MacLow2004}, the diffuse phases of the interstellar medium \citep{Elmegreen2004}, and in the intracluster media of clusters of galaxies \citep{Fang2009}.
It is known to play a dominant role in these environments by influencing the morphology, mixing characteristics, statistical properties and thermal structure of these  gaseous flows.

The interstellar medium (ISM) consists of complex, chaotic, and filamentary structures, where turbulent motions interacting with shocks play a major role \citep{Klessen2014}.  
Radio scattering and scintillation measurements revealed density fluctuations of the ISM consistent with a turbulent Kolmogorov spectrum over a wide range of scales from $10^6$ to $10^{18}$m \citep{Armstrong1995}.
A thorough understanding of these density structures is of great importance, as it has a direct influence on quantities like, e.g. the star formation rate, star formation efficiency, initial mass function, and core mass function
 \citep{MacLow2004, Ballesteros-Paredes2007, McKee2007, Hennebelle2008, Hennebelle2009, Hennebelle2011}. Therefore, a comprehensive understanding of the mass spectrum advected by a turbulent medium is of major interest.

%\todocom{Paragrpah ueber Observations notwendig?}
%In observations, in contrast to the theoretical work, the density power spectrum played an important role for analysing the turbulent properties of the ISM from the early beginning and studied extensively
%\citep[see e.g.][]{Spangler1990, Armstrong1995, Deshpande2000, You2007, Dutta2013}. On small scales ($\la10^3$AU) the electron density spectrum of \citet{Armstrong1995} are in agreement with a Kolmogorov scaling exponent of $-5/3$, whereas their nonradio data of the ISM ($\la 1$pc) indicate a scaling exponent of $-2$.  \citet{Deshpande2000} analysed the opacity of cold HI gas in our Galaxy finding shallower spectra with a scaling exponent of $-0.75$ on scales $0.07$pc$< \ell < 3$pc. \citet{Padoan2004} studied power spectra of column densities of molecular clouds measured with the $J=1-0$ $^{13}\mathrm{CO}$ line and got scaling exponents of $-0.74 \pm 0.08$ for the Perseus and Taurus, and $-0.76 \pm 0.08$ for the Rosetta cloud. Whereas different studies collected in table 1 of \citet{Spangler1990} are all in between $-2$ and $-1.52$. 
%\todocom{Ueberleitung}

%It is well known that in supersonic turbulence the density field is log-normally distributed \citep[e.g.~][]{Passot1994,Vazquez1994} ranging
%over many orders of magnitude depending on the compressive r.m.s. Mach number $\mathcal{M}_c$ of the system \citep{Konstandin2012b}.
%Although \citet{} presented a physically motivated fitting function 
The theoretical treatment of turbulence by \citet{Kolmogorov1941a} was derived in the incompressible limit. Hence, the velocity field and its statistical properties were main focus in theoretical works, also in the supersonic, compressible context.
Only recently theorists started analysing the density power spectrum in simulations \citep[e.g.][]{Padoan2004, Kim2005, Beresnyak2005, Kowal2007a, Kritsuk2007, Lemaster2009, Schmidt2008, Federrath2009}. 
\citet{Kim2005} performed a systematic study with a set of simulations finding that the density power spectra get shallower with increasing Mach number.
\citet{Federrath2009} investigated the influence of the mode of the forcing routine on the scaling exponent finding that compressive forcing yields a steeper spectrum than the solenoidal driven case.
\citet{Padoan2004} carried out turbulent simulations with weak and strong magnetic field finding that the density power spectrum is shallower in the strong case.
\citet{Kowal2007a} confirmed this by studying a set of weakly compressible simulations with varying strength of the magnetic fields.

\citet{Saichev1996} developed a model for the density field advected by a more general version of the Burgers equation, describing the extreme limit of highly compressible gas or equivalently very large Mach number flows.
The "normal" Burgers equation \citep{Burgers1939} treats the flow as network of interacting shock fronts and therefore neglects the pressure forces \citep[for a detailed description of the Burgers equation in the context of the interstellar medium see e.g.][]{Klessen2014}.
The model of \citet{Saichev1996} permits an analytic treatment of the density field of  compressible gas with dissipative and dispersive effects as well as pressure forces.
\citet{Saichev1996} derived analytically that the density spectra follow a power law $\propto k^{-\zeta}$, with different exponents $\zeta$ depending on the relation between the pressure forces and the magnitude of the velocity fluctuations of the medium. They obtained a density power spectrum $\mathcal{D}(k) \propto k^{-2}$
for strong pressure forces (i.e.~weak shocks) and  $\mathcal{D}(k) \propto k^{0}$ in the limit of negligible pressure forces (i.e.~strong shocks). Interestingly, these are also the results for a density distribution  consisting of a single discontinuity or an infinitely compressed peak, i.e.~a step function and a delta function, respectively. Note that the power spectrum of a field is determined by its strongest singularity.
In general a discontinuity in the $(q-1)$th order derivative leads to a power spectrum of the form $ \propto k^{-2q}$ \citep{Bec2007}.
For small values of the Mach number \citet{Tatsumi1974} and \citet{Tokunaga1976} showed that one-dimensional compressible turbulence reduces to the solution of two Burgers equations for density fluctuations, which implies a density spectrum scaling $\propto k^{-2}$ in the low Mach number regime in agreement with the result of \citet{Saichev1996}.

We therefore analyse the influence of the rms Mach number $\mathcal{M}$ on the scaling exponent of the density spectrum
\begin{equation}
\mathcal{D} (k, \, \mathcal{M}) \propto k^{\zeta(\mathcal{M})}\,,
\label{eq:Theory}
\end{equation} 
 guided by the mentioned theories.
We note that \citet{George1984} predicted a \textit{pressure} power spectrum for the incompressible case scaling like $\propto k^{-7/3}$ using dimensional arguments and confirmed this studying an axis symmetric jet. \citet{Bayly1992} analysed  the formal asymptotic expansion of the compressible Navier-Stokes equations about a uniform state predicting a $\propto k^{-7/3}$ scaling for the density power spectrum for weakly compressible flows.
\citet{Sridhar1994} and \citet{Goldreich1995} developed a theory for mildly compressible MHD turbulence, where the density power spectrum developed a Kolmogorov like $k^{-5/3}$ scaling.

The above mentioned numerical studies focused on the impact of different physical processes on the scaling behaviour of the density power spectrum.
The influence of the chosen fitting range, the used regression method, and the resolution of the simulation were not or only briefly discussed.
In \citet{Konstandin2015a} we demonstrated that the resolution and the employed regression method can have a major effect on the inferred scaling exponents of the power spectrum.
We showed that the velocity power spectrum is not converged with resolution in the highly supersonic case and that the scaling exponents still change dramatically with a resolution of $1024^3$
such that these measurements and the inferred trends should not be over interpreted.
Beside the resolution, the width, and the position of the fitting range were key factors, as the analysed velocity spectra were curved instead of following a power law. In the study presented here, we therefore analyse the \textit{local scaling exponents} of the density spectra and measure the variation of the
power law parameters with the scale $k$. We also discuss the influence of the resolution in detail, before 
we describe the scaling exponent as a function of the Mach number $\zeta(\mathcal{M})$.

The paper is organised as follows: Section \ref{sec:simulations} presents the properties of our simulations and Section \ref{sec:methods} the used methods. In Section \ref{sec:results} we show our results and discuss them in Section \ref{sec:discussion} in the context of previous studies. The last Section \ref{sec:summary} summarises our findings.  
 
%###############################################################################
% Simulations & Methods
%###############################################################################
\section{Simulations}
\label{sec:simulations}
We use FLASH4 \citep{Fryxell2000, Dubey2008} with the HLL5R solver \citep{Waagan2011} to numerically integrate the continuity equation and the Euler equation with a stochastic forcing
term $\mathbf{F}$ per unit mass on a uniform three-dimensional grid,
\begin{equation}
 \frac{\partial \rho}{\partial t} + \nabla (\rho \cdot \textbf{v} )=0 \,,
\label{eq:continuity}
\end{equation}
\begin{equation}
\frac{\partial \textbf{v}}{\partial t} +(\textbf{v} \cdot \nabla)\textbf{v}=-\frac{ \nabla p}{\rho} + \textbf{F} \, ,
\label{eq:Euler}
\end{equation}
where $\rho$ is the mass density, $\textbf{v}$ the velocity
field, and $p$ the pressure.
We simulate an isothermal medium throughout this study
such that $p=\rho\mathrm{c_{\mathrm{s}}}^2$, with the sound speed $\mathrm{c_{\mathrm{s}}}$.
This is a reasonable assumption due to efficient cooling processes observed in the dense
interstellar medium and molecular clouds \citep{Elmegreen2004}.
To measure the influence of the resolution on the results, we run simulations with $256^3$, $512^3$, and $1024^3$ grid cells.

We compute the random forcing field $\textbf{F}$ in Fourier space with a natural mix of rotational and compressive (solenoidal and compressive) modes \citep{Schmidt2006, Federrath2010, Konstandin2012b}.
The forcing only occurs on the large (integral) scales $1 \leqslant
\left|\textbf{k}\right| \leqslant 2$, measuring $k$ in units of
$2\pi/L$ with $L$ being the box size. The autocorrelation time-scale of the forcing algorithm is
set equal to $T=L/(\mathrm{c_{\mathrm{s}}}\mathcal{M})$, where $\mathcal{M}$ is the time averaged root mean
square Mach number. We fix the energy input rate to yield a constant energy flux through the scales. With this forcing scheme we drive rms Mach numbers of $\mathcal{M} \approx {1,5,10,16}$ in the statistically steady state of fully developed turbulence for the $256^3$, $512^3$, and $1024^3$ resolutions. Additionally, we performed $20$ simulations with systematically increasing Mach numbers ranging from $\mathcal{M}=0.65$ up to $\mathcal{M}=16$ equally spaced with $\Delta \mathcal{M}\approx 0.8$ for the lower resolution $256^3$. Therefore, we end up with $32$ different simulations in total.
We start with homogeneously distributed gas at rest and let it evolve for $\gtrsim 11\,T$ and store the relevant quantities every $0.1\,T$.

\begin{figure*}
\includegraphics[width=0.49\linewidth]{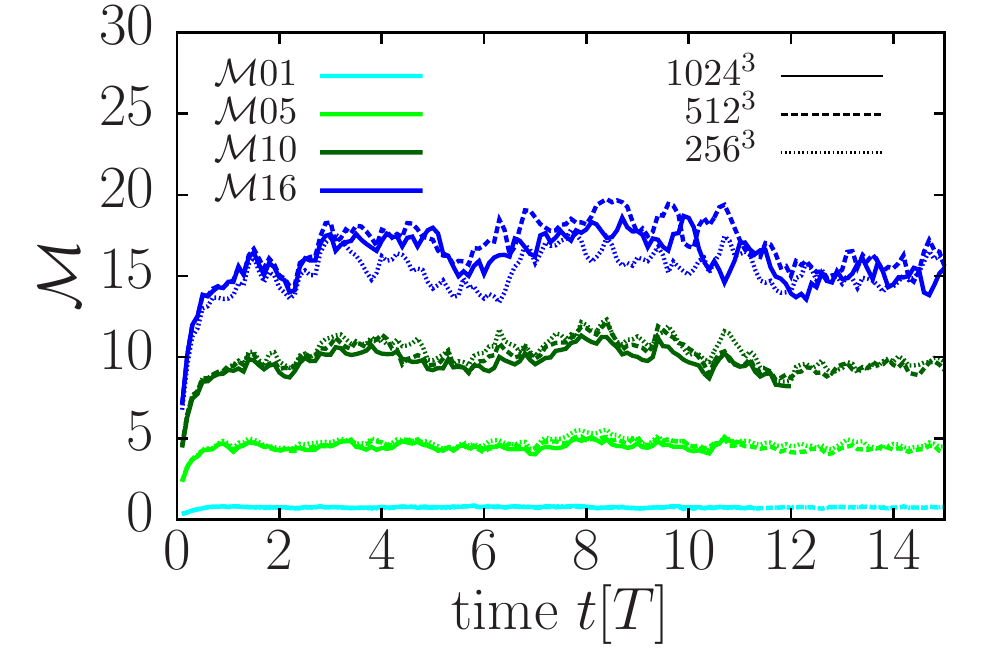}
\includegraphics[width=0.49\linewidth]{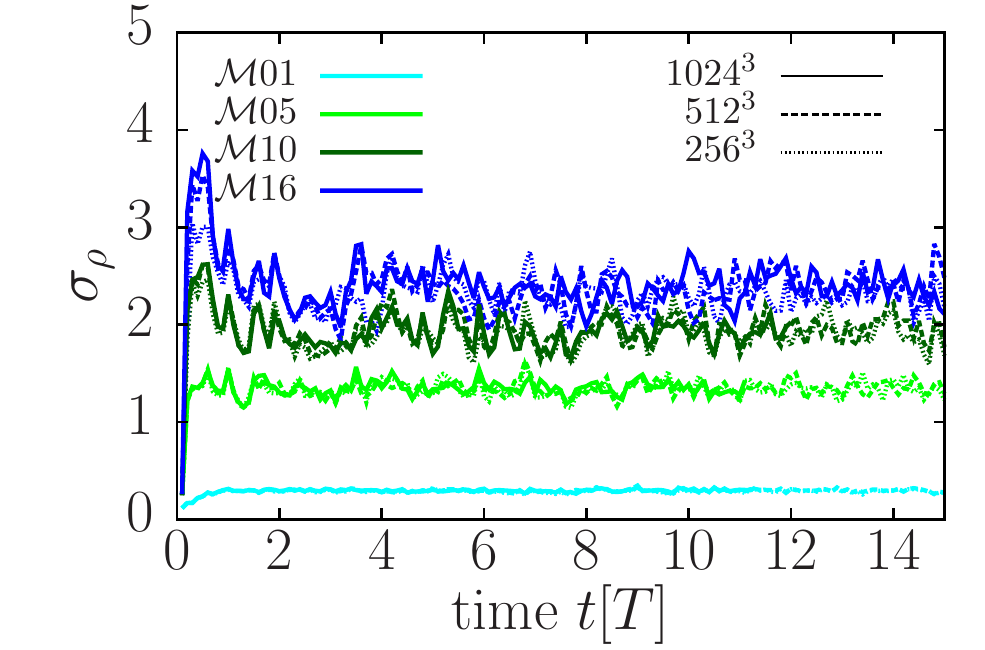}
\caption{Time evolution of the rms Mach number (left panel) and the standard deviation of the density $\sigma_{\rho}$ (right panel) at different resolutions $256^3$ (dotted), $512^3$ (dashed), and $1024^3$ (solid).}
\label{fig:time_evu}
\end{figure*}

Figure \ref{fig:time_evu} shows the time evolution of the rms Mach number (left panel) and the standard deviation of the density $\sigma_{\rho}$ (right panel) of the simulations for different resolutions.
After about two turbulent crossing times the velocity and the density fields are in the statistically steady state of fully developed turbulence. Both quantities measure the global one-point characteristic of the flow and are independent of the resolution.
We analyse in the following the spectral behaviour of the simulations for $t \geq 3T$.

\begin{figure*}
\includegraphics[width=0.48\linewidth]{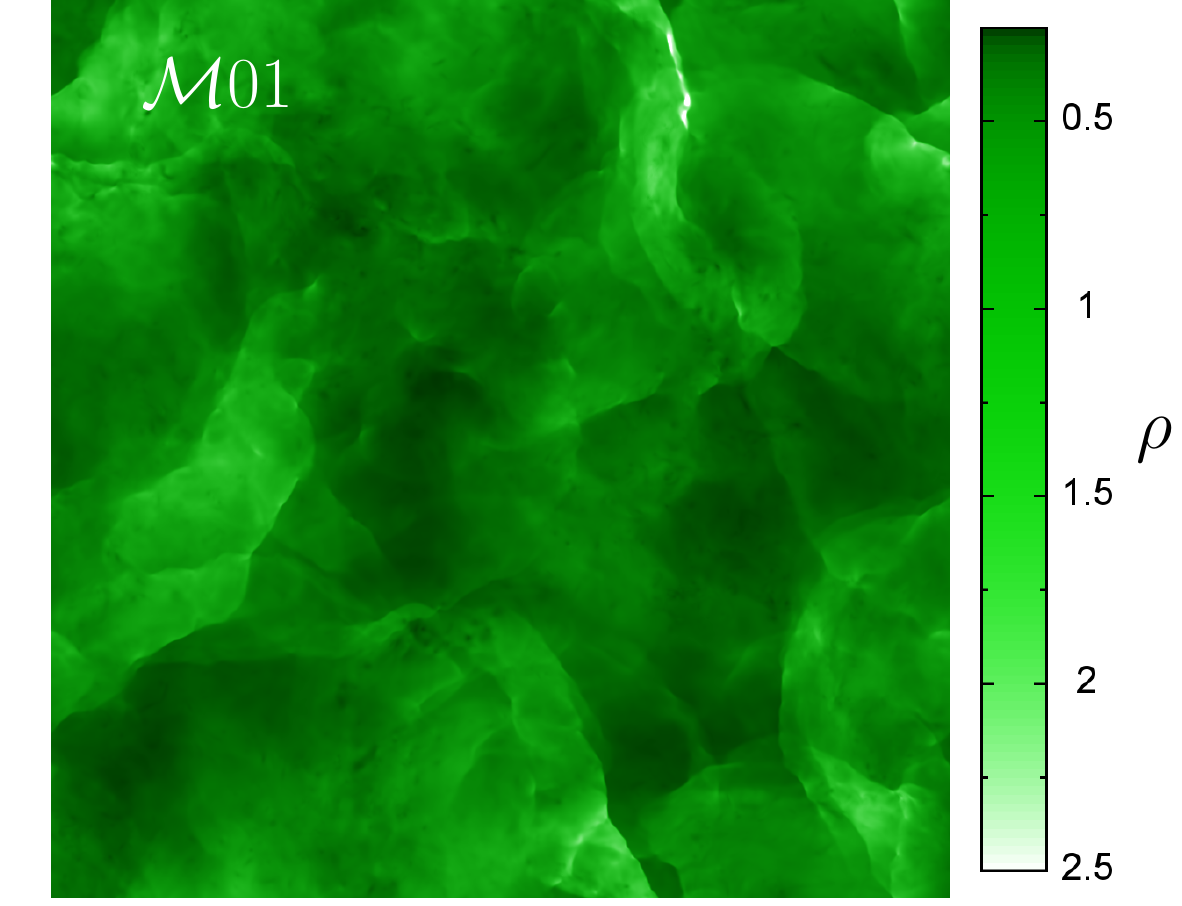}
\includegraphics[width=0.48\linewidth]{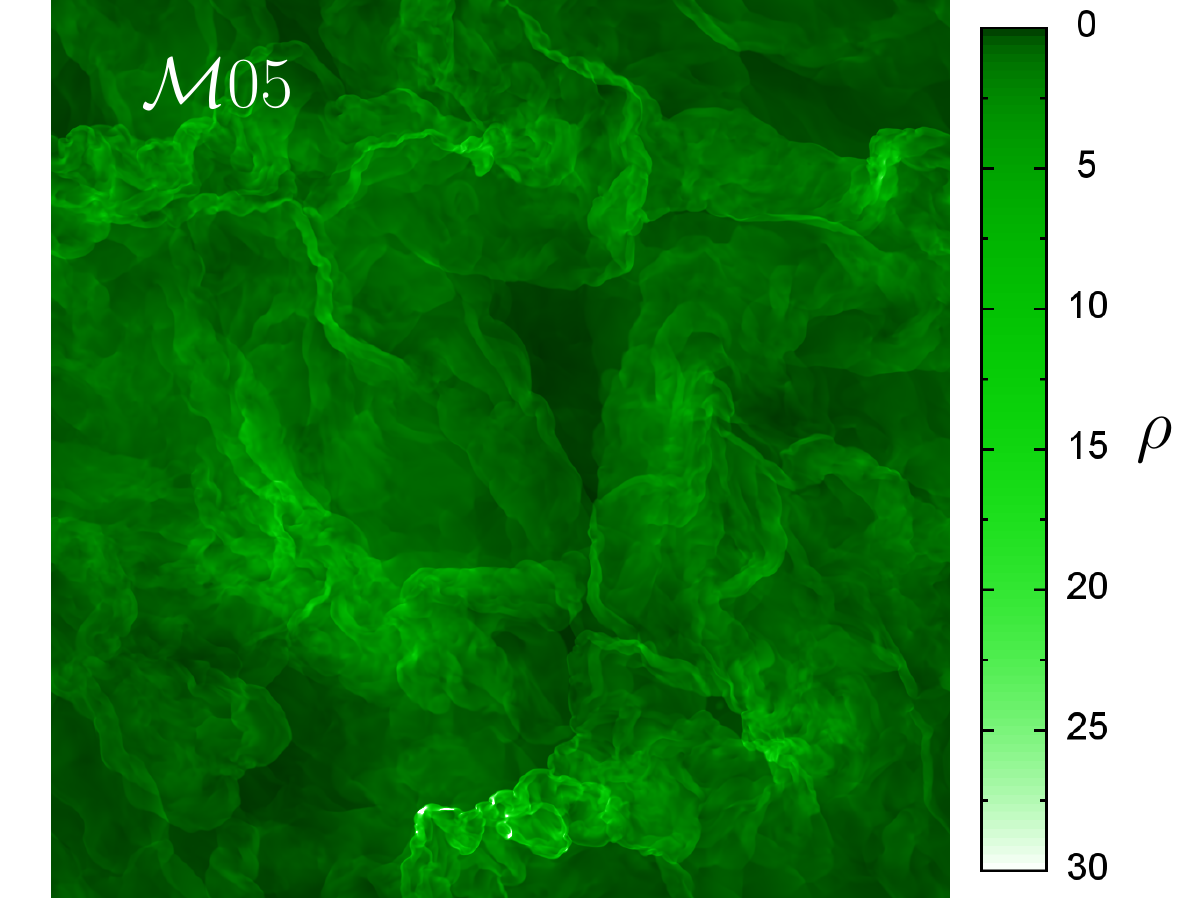}
\includegraphics[width=0.48\linewidth]{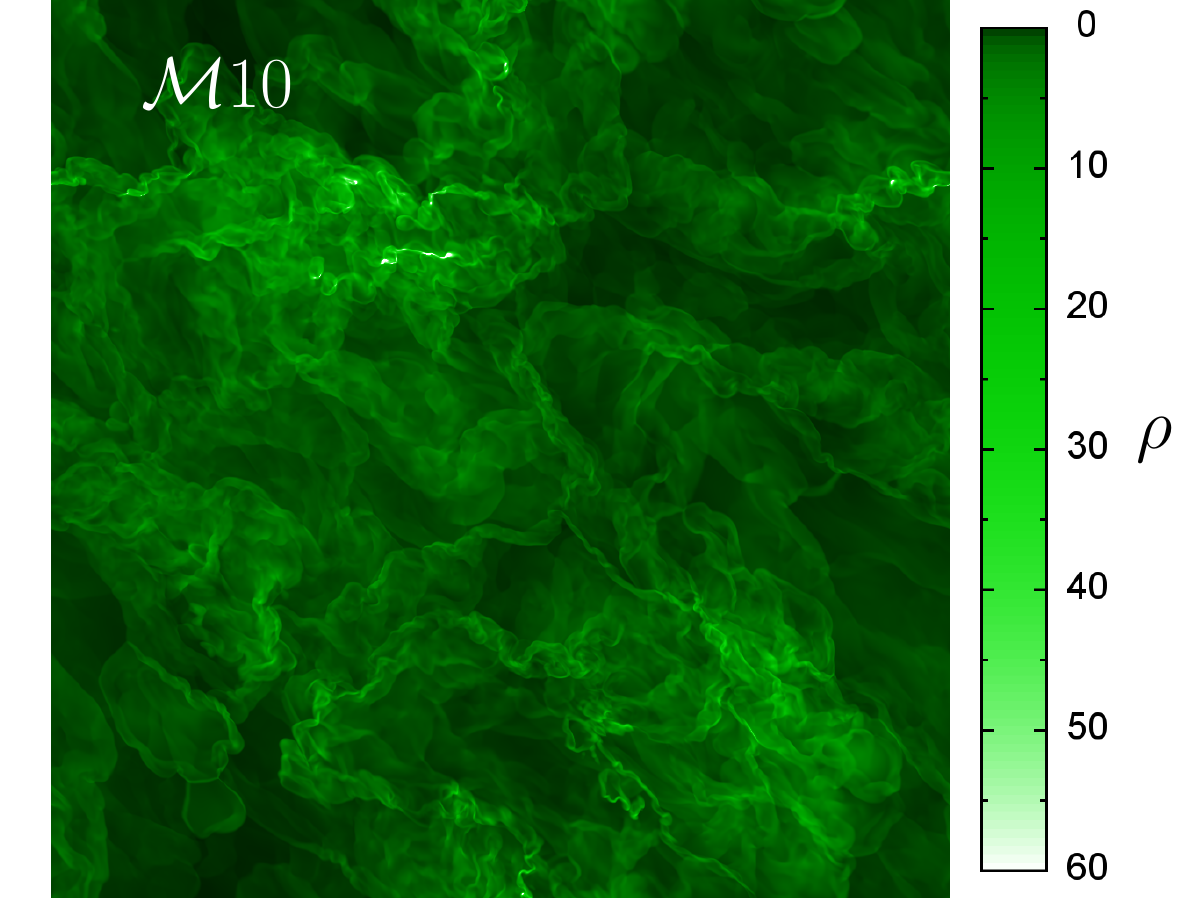}
\includegraphics[width=0.48\linewidth]{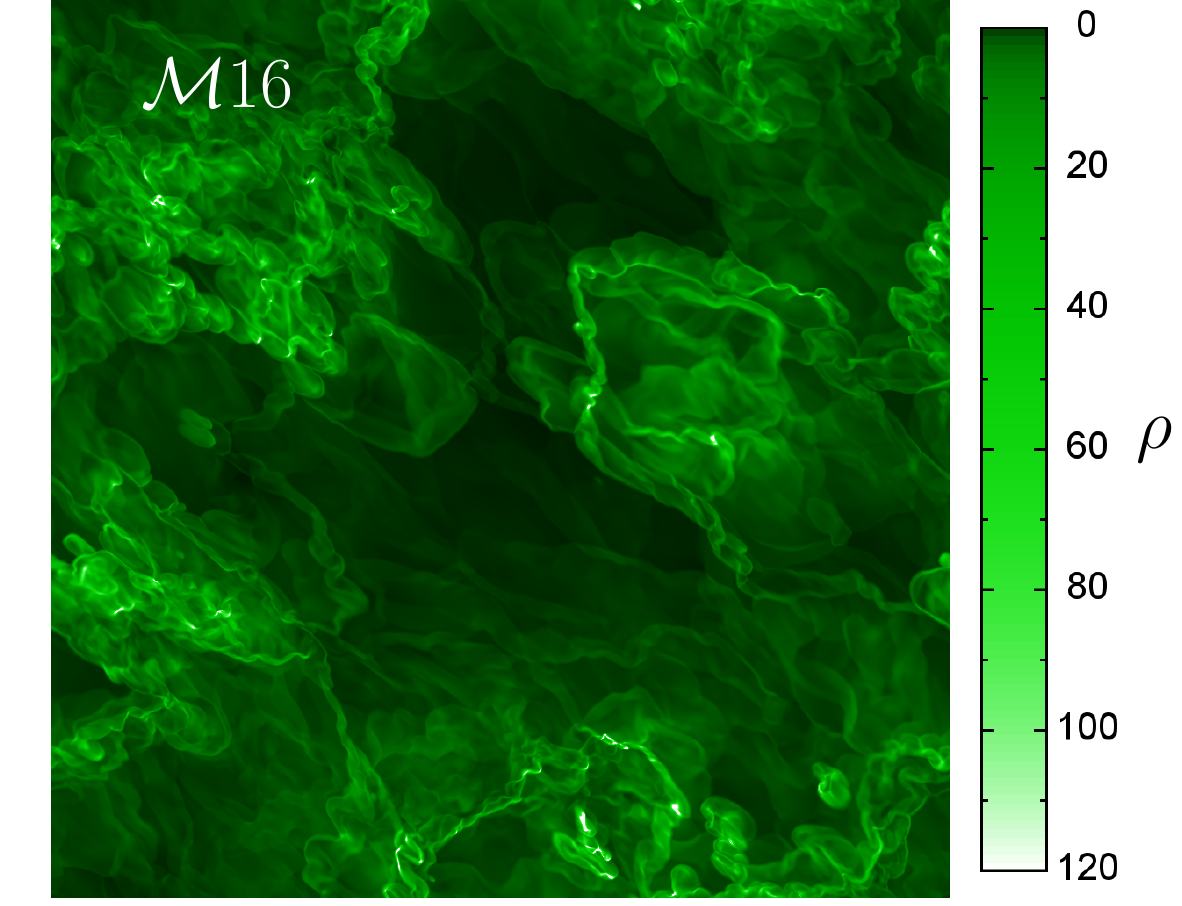}
\caption{The mass density in a cut trough the $xy$-plane at $z=0$ of the simulation
 $\mathcal{M}01$-$1024^3$ (top left), $\mathcal{M}05$-$1024^3$ (top right), $\mathcal{M}10$-$1024^3$ (bottom left), and $\mathcal{M}16$-$1024^3$ (bottom right) in the statistically steady state of fully developed turbulence.}
\label{fig:Slice}
\end{figure*}

Figure \ref{fig:Slice} presents the mass density in a cut trough the $xy$-plane at $z=0$ of the simulation
 $\mathcal{M}01$-$1024^3$ (top left), $\mathcal{M}05$-$1024^3$ (top right), $\mathcal{M}10$-$1024^3$ (bottom left), and $\mathcal{M}16$-$1024^3$ (bottom right) to illustrate the flow pattern in the statistically steady state of fully developed turbulence. The figures illustrate that the density contrast is increasing with the Mach number in general. A closer look reveals that the $\mathcal{M}01$ simulation lacks density fluctuations on small scales, whereas the higher Mach numbers show density fluctuations on all scales. This already indicates that the density spectrum is steeper at low Mach numbers, which we will analyse in the following in more detail.  
%###############################################################################
%METHODS SECTION
%###############################################################################
\section{Methods}
\label{sec:methods}
The Fourier spectrum of the density field is defined as
\begin{equation}
 \mathcal{D}(k, \, t_i,\,\mathcal{M}) \mathrm{d}k =  4\pi k^2 \hat{\rho}(k, \, t_i,\,\mathcal{M}) \cdot \hat{\rho}^*(k, \, t_i,\,\mathcal{M})\,\mathrm{d}k \, ,
\label{eq:def_spectrum}
\end{equation}
where $\hat{\rho}$ is the Fourier-transformed density field and $\hat{\rho}^*$ its complex conjugate.
We use the model described in \citet{Konstandin2015a} performing a hierarchical Bayesian linear regression on the logarithm of the power spectra.
We assume that every time snapshot $t_i$ of the density spectrum $\mathcal{D}(k, t_i,\,\mathcal{M})$ follows a power law,
\begin{equation}
\mathcal{D}(k,\, t_i,\,\mathcal{M}) = A(t_i,\,\mathcal{M}) k^{\zeta(t_i,\,\mathcal{M})} \sigma_{\Delta}(k,\,\mathcal{M}, \,t_i)\,,
\label{eq:Power_law}
\end{equation}
with the amplitudes $A(t_i,\,\mathcal{M})$, the scaling exponents $\zeta(t_i,\,\mathcal{M})$, and  the scatter term $\sigma_{\Delta}(k,\,\mathcal{M}, t_i)$.
This method has the advantage that errors and uncertainties are treated self consistently, averaging the data is not necessary, as all snapshots of the spectra are fitted simultaneously, and it provides valid estimates for the fitting parameters, their errors, as well as their time variation.
We apply the method to extended fitting ranges, as well as small fitting windows only containing seven data points $[k-3, k+3]$, which we move systematically over the spectrum and interpret the result as \textit{local scaling exponent} at the scale $k$.
We refer the reader to \citet{Konstandin2015a} for a detailed description of the Bayesian model, various test on the parameter estimates (like the influence of the fitting range), and a comparison with ordinary linear regression methods
applied to the averaged spectrum as well as the individual power spectra. We focus the discussion in the following on the mean group slope, its uncertainty and  variation with time. The mean group slope can be interpreted as a time averaged scaling exponent.

%###############################################################################
%RESULTS SECTION
%###############################################################################
\section{Results}
\label{sec:results}
The Parseval theorem 
\begin{equation}
\sigma_{\rho}^2 + \mu_{\rho}^2 = \int_0^{\infty} \mathcal{D}(k)\, \mathrm{d}k
\label{eq:Parseval}
\end{equation}
links the integral of the density spectrum with the variance and the squared mean of the mass density probability distribution function.
Hence, the first two moments of the density distribution describes the area below the density power spectrum.
We therefore discuss in Section (\ref{sec:DensityPDF}) the probability density function with its moments before we focus on the density spectrum and its scaling behaviour in the following Sections.
\begin{figure}
\includegraphics[width=0.98\linewidth]{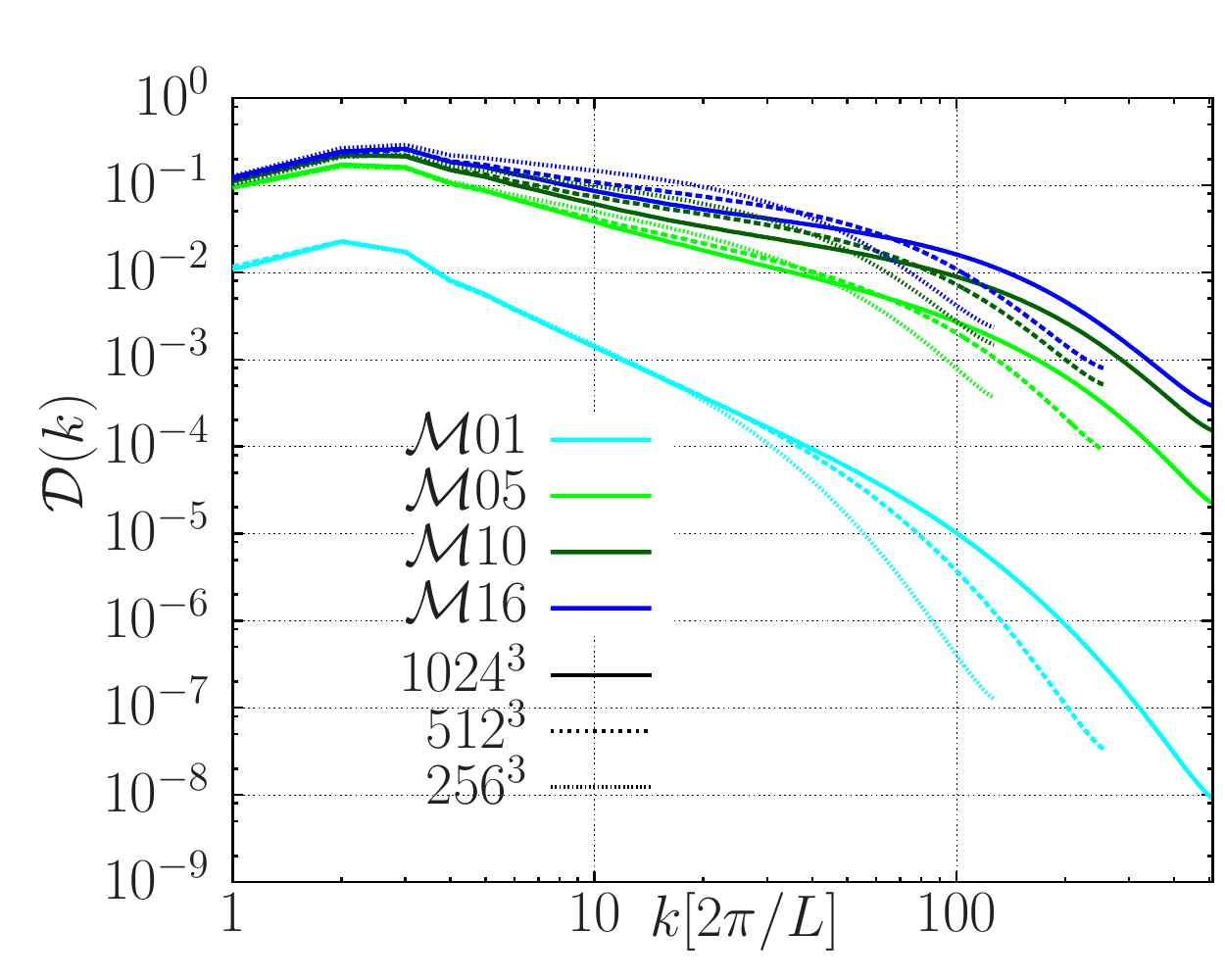}
\caption{Density power spectra of the simulation with $256^3$ (dotted), $512^3$ (dashed), and $1024^3$ (solid) resolution and different Mach numbers averaged over time $t \geq 3\,T$ in the statistically steady state of fully developed turbulence. }
\label{fig:Spec_dens}
\end{figure}
Figure \ref{fig:Spec_dens} presents the density power spectra for different Mach numbers and different resolutions. The mildly supersonic simulation $\mathcal{M}01$ stands out because of its lower amplitude in comparison to the highly supersonic simulations.
Figure \ref{fig:Spec_dens} indicates that the area gain of the density spectrum for higher Mach numbers happens through a shallower scaling exponent instead of an increasing amplitude. Comparing the resolutions in Figure \ref{fig:Spec_dens} reveals that the  spectra with $\mathcal{M}01$ are hardly distinguishable for $k<20$, whereas in the highly supersonic simulations the resolution has an influence for $k \gtrsim 5$, depending on the Mach number.

%###############################################################################
%PDF SUBSECTION
%###############################################################################
\subsection{The probability distribution function (PDF) of the mass density}
\label{sec:DensityPDF}
\citet{Passot1998} propose a heuristic model for the density PDF based on the density contrast from the shock jump condition in an isothermal medium. They assume that the logarithmic density variation $\Delta s$ by the shocks can be interpreted as an additive random process changing the logarithmic density $s=\log{(\rho)}$. Arguing with the central limit theorem they propose a normal distribution for the logarithmic density $s$ and a log-normal distribution for the density. The moments of these distributions are connected via
\begin{equation}
\mu_{\rho} =\exp{(\mu_{s} + \sigma_s^2/2)} 
\end{equation}
for the mean and 
\begin{equation}
\sigma_{\rho}^2 = \mu_{\rho}^2 (\exp{(\sigma_s^2)} - 1)
\end{equation}
for the standard deviation.
We choose the parameters of the simulations in this way that the box size $L=1$, the total volume $V=1$ and the total mass $M=1$ such that
$\mu_{\rho} =1$ implyies
\begin{equation}
\mu_s = -\sigma_s^2/2
\label{eq:LN_mean}
\end{equation}
and 
\begin{equation}
\sigma_{\rho}^2 =\exp{(\sigma_s^2)} - 1\,.% = \mathrm{b}^2\mathcal{M}^2\,.
\label{eq:LN_stddev}
\end{equation}

\citet{Hopkins2013b} developed a model for the density distribution taking mass conservation and intermittent fluctuations into account with a non log-normal shape in order to explain the deviations from the relations between the moments seen in 3-dimensional numerical simulations \citep[e.g.][]{Kowal2007a, Kritsuk2007, Schmidt2009, Price2010, Federrath2010, Konstandin2012b, Molina2012}.
In this model the deviations from the log-normal shape can be expressed by a single parameter $T$
%\footnote{In this paragraph $T$ refers to the parameter in the \citet{Hopkins2013b} model. In the rest of the paper $T$ is the timescale of the forcing routine introduced in Section \ref{sec:simulations}.}
, with $T=0$ for a log-normal distribution.

\begin{figure}
\includegraphics[width=0.95\linewidth]{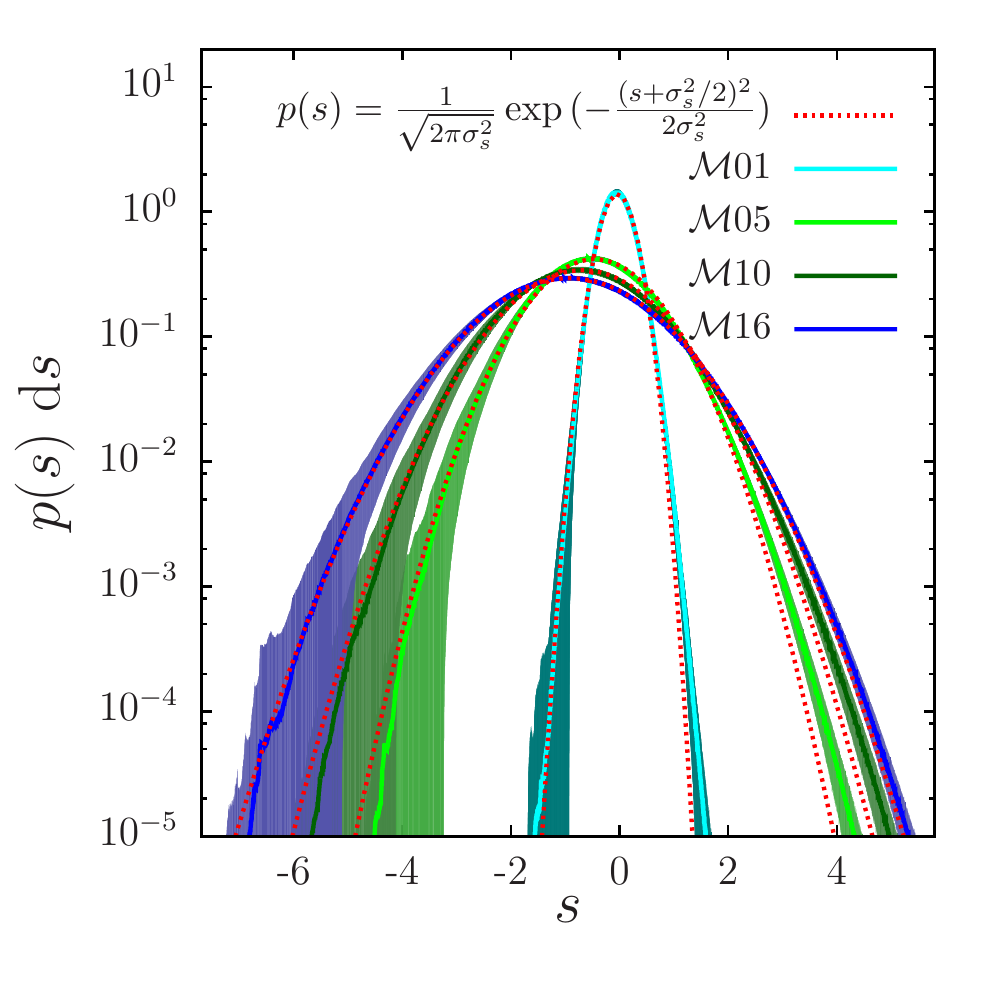}
\caption{Probability distribution function of $s$ measured in the simulations with $1024^3$ resolution and at different Mach numbers. The shaded areas indicate the one sigma time variations of the PDFs. The red dotted lines correspond to a normal distribution with only one parameter, as we expressed the mean value via $\mu_s = -\sigma_s^2/2$.}
\label{fig:PDF}
\end{figure}

\begin{figure}
\includegraphics[width=0.95\linewidth]{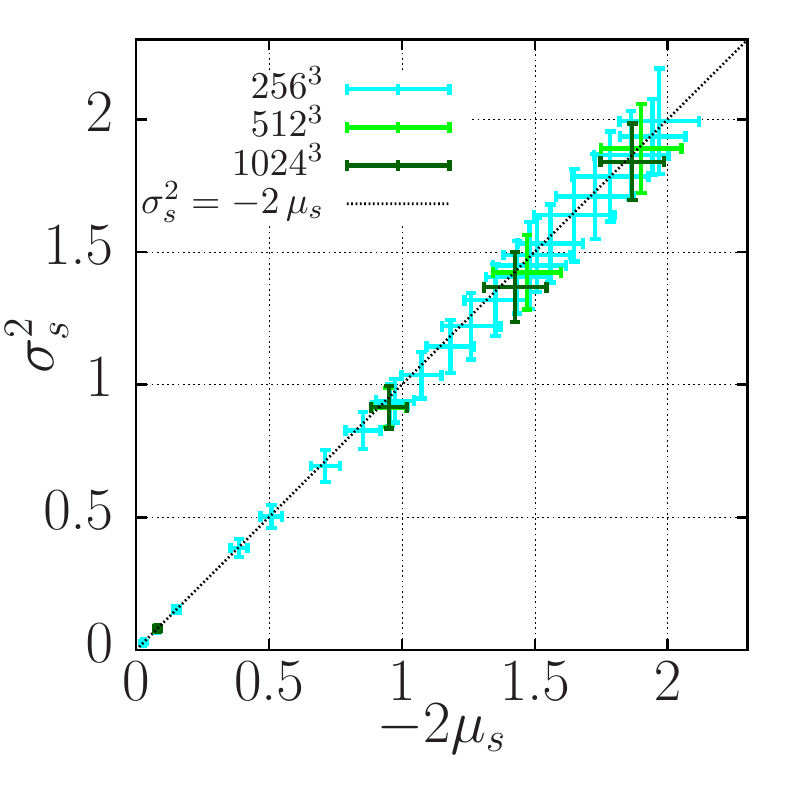}
\includegraphics[width=0.95\linewidth]{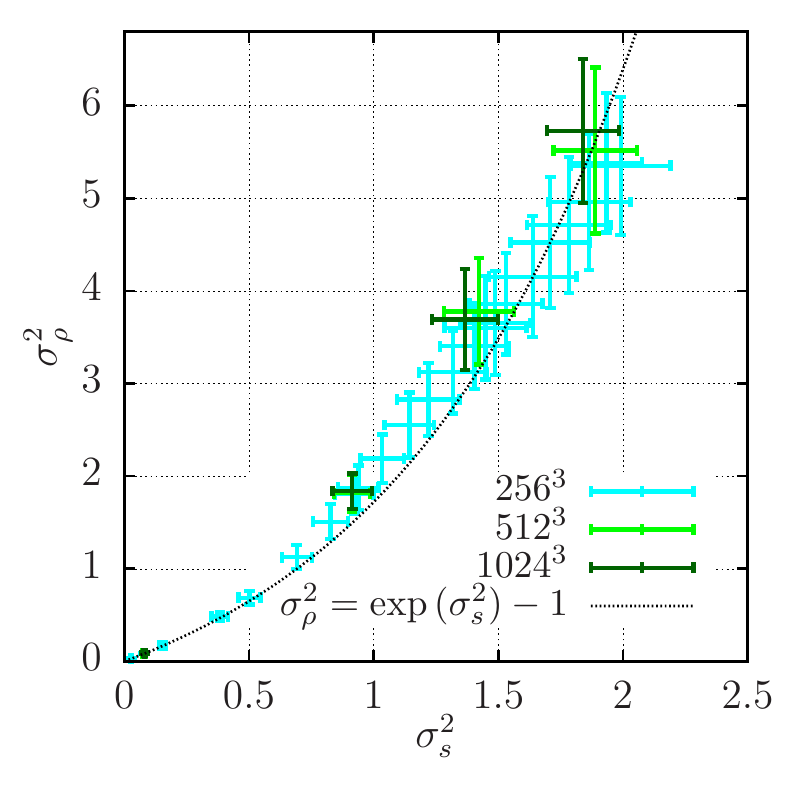}
\caption{Relation between the moments of the density field assuming a log-normal distribution. Top panel: Mean of the logarithmic density $s$ versus its squared standard deviation together with equation (\ref{eq:LN_mean}). Bottom panel: Squared standard deviations of the density $\rho$ and the logarithmic density $s$ together with equation (\ref{eq:LN_stddev}).}
\label{fig:LNMoments}
\end{figure}

Figure \ref{fig:PDF} shows the PDF of $s$ measured in the simulations with $1024^3$ resolution and at different Mach numbers together with
normal distributions depending only on one parameter $\sigma_s$. We express the mean of the
normal distributions with equation (\ref{eq:LN_mean}). The normal distributions are in excellent agreement with the measured PDFs also in the wings of the distributions.
We show the relations (\ref{eq:LN_mean}) and (\ref{eq:LN_stddev}) between the moments in Figure \ref{fig:LNMoments} to quantify the discrepancy between the log-normal assumption and our simulations.
Our data show only negligible deviations from both relations
indicating that our density PDFs are closer to a log-normal shape than the ones analysed by
\citet{Hopkins2013b}. Therefore, we find a $T$ parameter scattering around zero (e.g.~$T=-0.0003 \pm 0.0003$ in the simulation $\mathcal{M}15$-$1024^3$) for all Mach numbers, which we calculated with the relation  between the moments and the $T$ parameter given in equation (6) of \citet{Hopkins2013b}. The main difference between the simulations presented here and these in \citet{Federrath2008b} and \citet{Konstandin2012b} is the decomposition of the forcing scheme. We use a here a forcing field, which contains both solenoidal (transverse) as well as compressive (longitudinal) modes in equal parts, whereas the above mentioned studies focus on the the extreme cases of purely solenoidal and purely compressive forcing.

%###############################################################################
%Sigma-M SUBSECTION
%###############################################################################
\subsection{The $\sigma_s$-$\mathcal{M}$ relation}
In the next step we analyse the relation between the standard deviation of the density distribution and the Mach number of the turbulent flow.
\citet{Passot1998} concluded from the shock jump condition and the central limit theorem that
the standard deviation of the logarithmic density $\sigma_s \approx \mathrm{b} \mathcal{M}$ with the proportionality constant $\mathrm{b}$.
This leads to $\sigma_{\rho} \approx \sigma_{s} \approx \mathrm{b}\mathcal{M}$ when Taylor expanding equation (\ref{eq:LN_stddev}) to first order. \citet{Federrath2008b} analysed the influence of different forcing fields on the relation
\begin{equation} 
\sigma_s^2 = \log{(1+\mathrm{b}^2\mathcal{M}^2)}\,.
\label{eq:Passot_stddev}
\end{equation}
They assume a linear relation between the standard deviation of the density field and the Mach number \citep{Padoan1997} and use equation (\ref{eq:LN_stddev}) to express the standard deviation of the logarithmic density.
They found $0 \leq \mathrm{b}\leq 1$
depending on the decomposition of the forcing with $\mathrm{b}=1/3$ for purely solenoidal forcing and $\mathrm{b}=1$ for purely compressive. This parameter $\mathrm{b}$ can be interpreted as the ratio between the compressive Mach number over the total rms Mach number of the flow \citep{Konstandin2012b}
\begin{equation}
\mathrm{b}=\frac{\mathcal{M}_c}{\mathcal{M}}\,.
\end{equation} 
The longitudinal modes of the velocity field contain colliding and dispersing flows, which cause density fluctuations, whereas the transverse modes have no influence.

The jump condition $\rho(t_{i+1})=\rho(t_i)\mathcal{M}^2$ indicates that we have to express the 
PDF at the time $t_{i+1}$ conditioned on the past time $\mathrm{p}(\rho( t_{i+1}) | \, \rho(t_i))$.
A stochastic processes is said to be Markovian, if the system only depends on the state of the previous time step, but not on those before. 
As the shock jump condition suggests this behaviour, we will use the Fokker-Planck transport equation for random variables,
\begin{equation}
\frac{\partial}{\partial t} p(s;\,t) =-\frac{\partial}{\partial s} [B_1(s;\,t) p(s;\,t))]+\frac{1}{2}\frac{\partial^2}{\partial s^2} [ B_2^2(s;\,t) p(s;\,t)]\,.
\label{eq:Fokker}
\end{equation}
with the drift parameter $B_1(s,\,t)$ and the diffusion coefficient $B_2(s,\,t)$ to determine the steady state of the PDF of $s$ 
\citep{Risken1996, Baudoin2014}.
An equivalent description of the process can be achieved by expressing the random variable itself in a stochastic differential equation
instead of the time evolution of its PDF
\begin{equation}
\mathrm{d}s = B_1(s,\,t) \mathrm{d}t + B_2(s,\,t)\mathrm{d}W\,,
\label{eq:Diffusion}
\end{equation}
with the random process $\mathrm{d}W$.
Following the idea of \citet{Passot1998} and \citet{Federrath2008b} we assume $B_1=0$ and $B_2^2=\log{(1+\mathrm{b}^2\mathcal{M}^2)}$ in equation (\ref{eq:Diffusion}), which describes the density fluctuations as random process caused by the shocks. With this ansatz the Fokker-Planck equation has no stationary distribution. Instead we get with the initial values $t_0$ and $s_0$ and the initial condition
$p(s;\,t_0|s_0;\,t_0) = \delta(s - s_0)$ 
\begin{equation}
p(s;\,t | s_0;\, t_0) = \frac{1}{\sqrt{2\pi(t-t_0)}\sigma_s} \exp{\left(-\frac{(s-s_0)^2} {2 \sigma_s^2 (t-t_0) }\right)}\,,
\end{equation}
which is the dispersion relation of a Brownian motion. This is expected as the model does not contain the redistribution of the mass according to the hydrodynamical equations. In the next step we assume that the pressure difference between the ambient/average pressure $\mu_p$ and the pressure $p$ of the different positions of the medium relaxes the gas compressed by the shocks. We express the pressure term $\rho^{-1}\boldsymbol{\nabla} p =c_s^2\boldsymbol{\nabla}s \approx c_s^2 (s - \mu_s) $ and make the ansatz 
\begin{equation}
B_1(s(t)) =-1/\tau_{\alpha}(s-\mu_s)\quad B_2^2=1/\tau_{\beta} \log{(1+\mathrm{b}^2\mathcal{M}^2)}\,,
\label{eq:Ansatz}
\end{equation}
with the efficiency parameter $1/\tau_{\alpha}$ and $1/\tau_{\beta}$ for the different processes, which will be determined later. 
Equation (\ref{eq:Diffusion}) reads then
\begin{equation}
\mathrm{d}s(t) = -(s-\mu_s)/\tau_{\alpha} \mathrm{d}t + \sqrt{( \log{(1+\mathrm{b}^2\mathcal{M}^2)}/\tau_{\beta})} \mathrm{d}W\,,
\end{equation}
which is a Ornstein-Uhlenbeck process \citep{Pope2000, Baudoin2014} describing the non equilibrium evolution of the density field.

\begin{figure}
\includegraphics[width=0.95\linewidth]{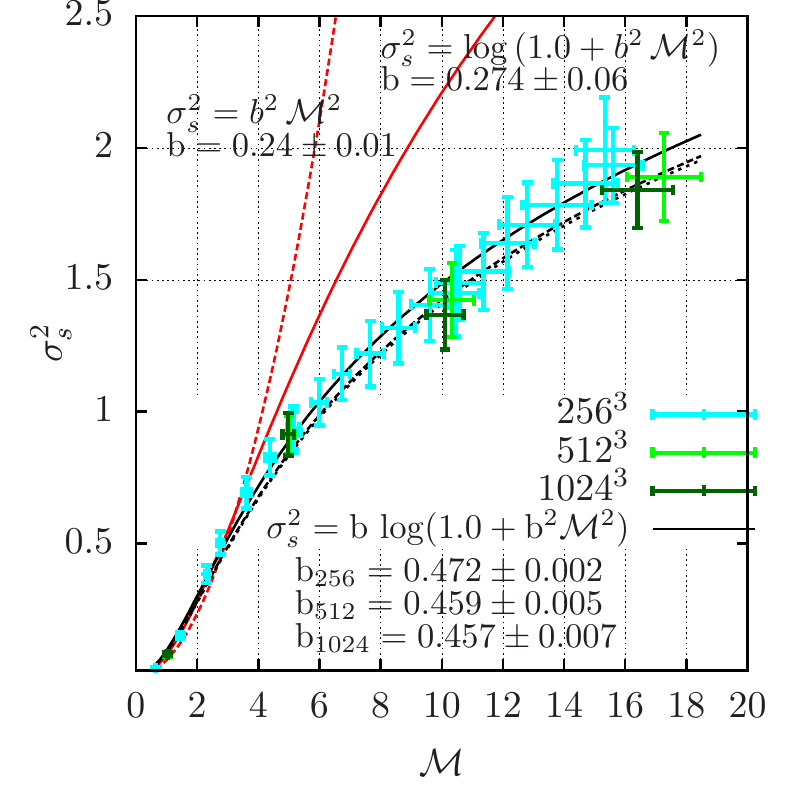}
\caption{Squared standard deviation of the mass density $\sigma_{s}^2$ as a function of the Mach number.
We show the relation between the Mach number and the standard deviation of the logarithmic density $\sigma_{s}^2 = \mathrm{b}^2\mathcal{M}^2$ (red dashed line),
$\sigma_{s}^2 = \log{(1+ \mathrm{b}^2\mathcal{M}^2)}$ (red solid line) and the new proposed formula  $\sigma_{s}^2 = \mathrm{b} \log{(1+ \mathrm{b}^2\mathcal{M}^2)}$ (black lines).
The fits of the data with $256^3,\,512^3,\,1024^3$ resolution are solid, dashed, dotted, respectively.}
\label{fig:Mach_vs_rho}
\end{figure}

The stationary solution with ansatz (\ref{eq:Ansatz}) can be calculated with the Fokker-Planck equation (\ref{eq:Fokker}), which gives
\begin{equation}
p(s,\,t | s_0,\,t_0) =  \mathcal{N}(\mu_s,\,\sigma_s^2)\,,
\end{equation}
a normal distribution with mean $\mu_s$ and squared standard deviation $\sigma_s^2=\tau_{\alpha}/2\tau_{\beta} \log{(1+\mathrm{b}^2\mathcal{M}^2)}$ in agreement with
the theories of \citet{Passot1998} and \citet{Federrath2008b} besides the prefactor $\tau_{\alpha}/2\tau_{\beta}$. The timescale $\tau_{\alpha}$ associated with the pressure term can be interpreted as the time for redistributing shocked and diluted gas and $\tau_{\beta}$ is the timescale associated with the shocks occurring during the time $\mathrm{d}t$. We express the timescale of the pressure term with  the dynamical timescale $\tau_{\alpha}=L/2 \mathcal{M}$. Whereas we estimate the shock frequency with two times the compressive Mach number $\tau_{\beta}=L/4 \mathcal{M}_c$ as only the longitudinal part of the velocity field contributes to advecting flows, however it counts twice for opposed shock fronts. Therefore we end up with $\tau_{\alpha}/2\tau_{\beta} = \mathrm{b}$ such that 
\begin{equation}
\sigma_s^2 = \mathrm{b}\log{(1+\mathrm{b}^2\mathcal{M}^2)}\,.
\label{eq:new}
\end{equation}
Figure \ref{fig:Mach_vs_rho} presents $\sigma_s^2$ as a function of the Mach number $\mathcal{M}$ for the simulations at different resolutions.
Also shown are the relations proposed by \citet{Passot1998} and \citet{Federrath2008b}, both fitting the data for $\mathcal{M}<4$, but significantly overestimate $\sigma_s^2$ in the high Mach number regime.
We measure with these models $\mathrm{b} = 0.24 \pm 0.01$ ($\sigma_s = \mathrm{b}\mathcal{M}$)
and $\mathrm{b} = 0.27 \pm 0.06$ ($\sigma_s^2 = \log{(1+\mathrm{b}^2\mathcal{M}^2)}$) limiting the fitting range to $\mathcal{M}<4$.
These measurements are slightly smaller than $\mathrm{b}=1/3$ proposed for purely solenoidal forcing.
In contrast the fit of relation (\ref{eq:new}) is in agreement with the data
for all Mach numbers. We measure $\mathrm{b}=0,472 \pm 0.002,\, 0.459 \pm 0.005,\, 0,457 \pm 0.007$ for 
$256^3,\, 512^3,\, 1024^3$ resolution respectively. The new proposed model could also explain the measurements of \citet{Konstandin2012b},
who found strong deviations from equation (\ref{eq:Passot_stddev}) for purely solenoidal forcing ($\mathrm{b}\approx1/3$), whereas no discrepancies where found for purely compressive forcing ($\mathrm{b} \approx 1$).
We will analyse the influence of the forcing field on this relation in future work to test this hypothesis.

%###############################################################################
%Local scaling 
%###############################################################################
\subsection{The local scaling exponents}
\label{sec:localSE}
\begin{figure}
\includegraphics[width=0.98\linewidth]{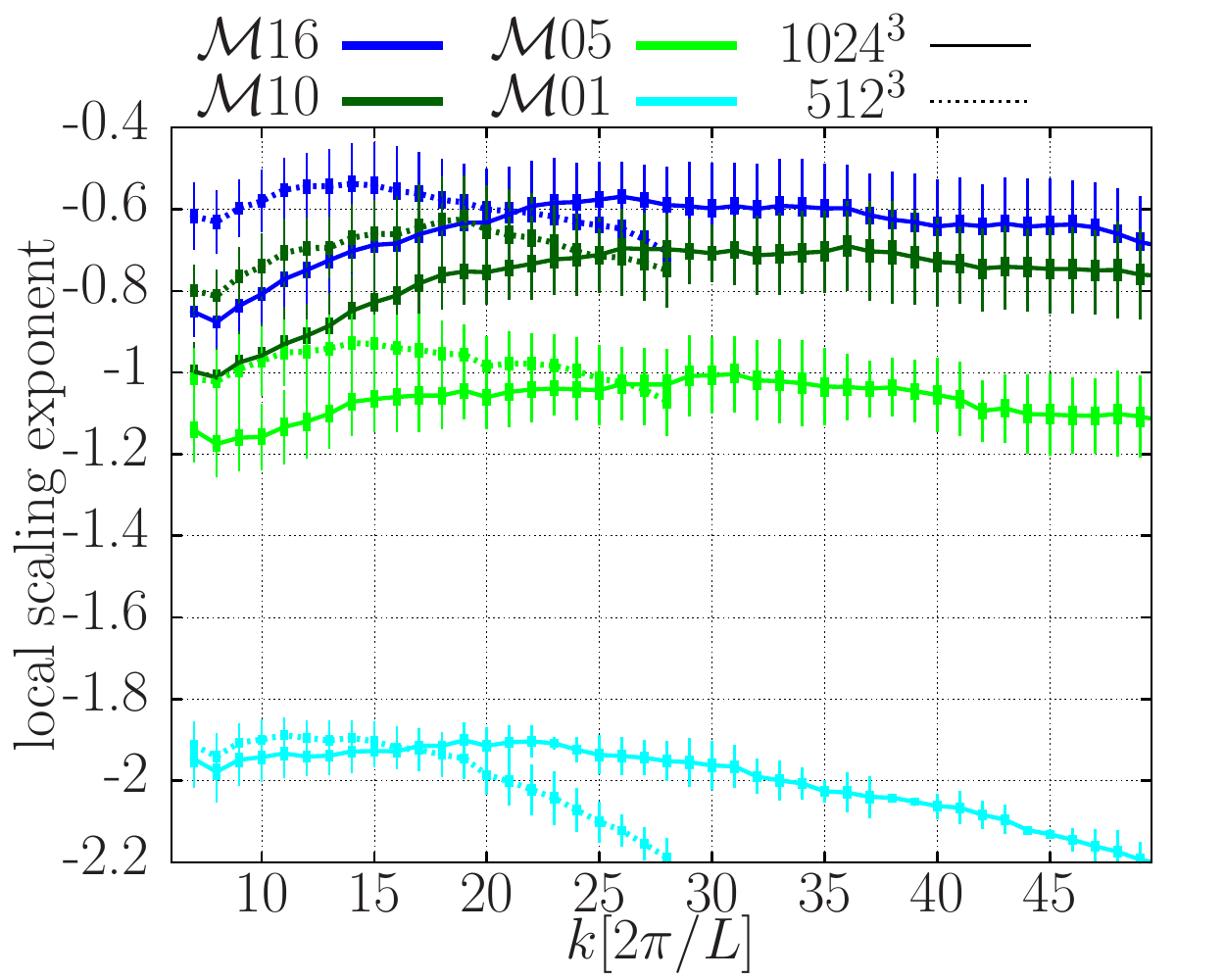}
\caption{The local scaling exponents of the density power spectra as a function of the wave number $k$ at $512^3$ (dashed) and $1024^3$ (solid) resolution and for different Mach numbers.
The local scaling exponents are the results of a Bayesian regression applied to systematically moved fitting windows $[k-3, k+3]$. The thick error bar indicates the uncertainty of the group scaling exponent and the thin error bar states the variation of the scaling exponent with time \citep[see][for more details]{Konstandin2015a}.}
\label{fig:Localslope_mix_dens}
\end{figure}
Figure \ref{fig:Localslope_mix_dens} shows the local scaling exponents of the density power spectra as a function of the scale for different Mach numbers. 
Recall that every point corresponds to a fit over a range of $\Delta k = 6$ such that the first points are the results of a fit $k \in [4:10]$ centred at $k_c=7$. 
The lines connect the mean population slope, the thick error bar corresponds to its $1\,\sigma$ uncertainty, and the thin error bar states the $1\,\sigma$ variation with time.
The dashed lines indicate the results of the lower resolution runs ($512^3$) and the solid lines are for the higher resolution ($1024^3$).
Note that simulations with resolution of $512^3$ and $1024^3$ the wave numbers above $k \gtrsim 16$ and
$k \gtrsim 32$ are known to be dominated by numerical effects \citep{Kritsuk2007, Federrath2010, Konstandin2012a, Bertram2015, Konstandin2015a}, respectively.
Therefore, we stop showing the scaling exponents for the $512^3$ resolution simulations for $k>28$ to keep the plot readable and focus the interpretation in the following discussion on wave numbers smaller than $k \lesssim 32$.
 
All local scaling exponents in Figure \ref{fig:Localslope_mix_dens} are nearly constant and only increase slightly with the scale $k$. This curvature increases with higher Mach numbers, such that the difference between the large scale $k=7$ and the intermediate scale $k\approx25$ local scaling exponent is  $\approx 0.3$ for the $\mathcal{M}16$-$1024$ simulation.
The low Mach number run $\mathcal{M}01$ has a small range ($k\in [7:18]$) in which both resolutions have comparable and nearly constant local scaling exponents.
Whereas, the higher Mach number runs show a resolution dependence of the local scaling exponents already on scales close to the forcing routine, where the scaling exponents of the higher resolution simulations are systematically smaller ($\approx 0.2$) than their lower resolution counterparts. 

Our measurements are in agreement with the results of \citet{Saichev1996} predicting a spectrum $\propto k^{-2}$ for weak shocks and $\propto k^0$ for infinitly strong shocks.
For a given resolution, all the curves of the local scaling exponent for flows with higher $\mathcal{M}$ lie systematically above the ones for lower Mach numbers, with values of about $-2$ for $\mathcal{M}=1$.
In the limit of weak shocks the density profile is composed of sawtooth or step functions. Contrary, in the limit of strong shocks the density profile is peaky with few spots containing most of the mass (i.e. delta functions, see Figure \ref{fig:Slice}).
Following the ideas of \citet{Kim2005} our results can be interpreted as follows.
The density jump in an isothermal shock is proportional to the square of the Mach number yielding very large density contrasts.
Therefore, conserving the total mass, the gas gets more concentrated in shocks with increasing Mach number causing a more peaky density distribution. 
This leads to a shallower density spectrum for the higher Mach numbers. 
Increasing the resolution has the opposite effect. If a simulation with high Mach number is poorly resolved most of the mass is contained in few grid cells creating a peaky density distribution with shallow scaling exponent. With increasing resolution substructures get refined and mass gets distributed over more grid cells causing a steeper density spectrum.
 
\begin{figure}
\includegraphics[width=0.98\linewidth]{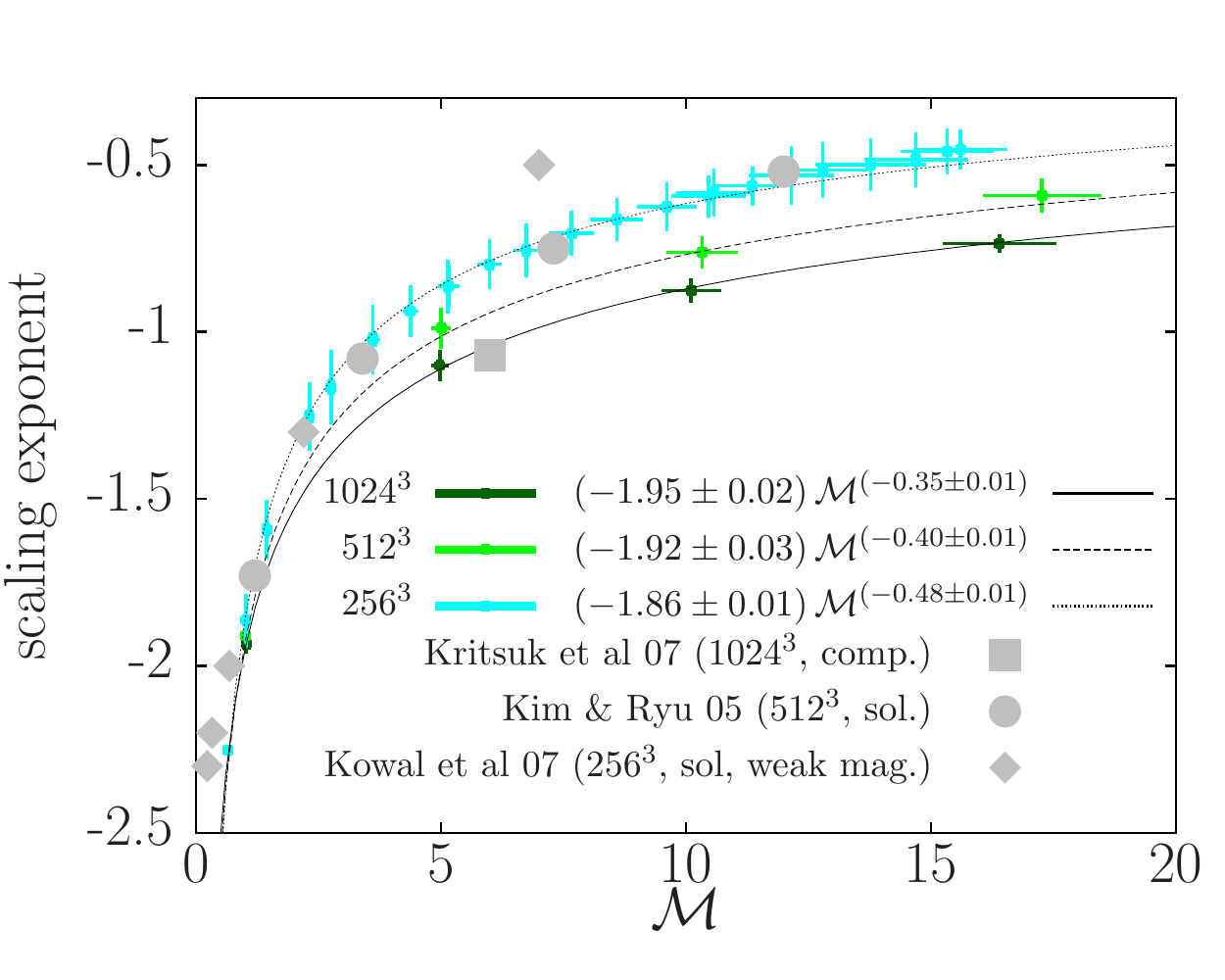}
\caption{Scaling  exponents as a function of the Mach number, measured in $k \in [4:10]$ for the simulations with $256^3$ (cyan), $k \in [4:17]$  for $512^3$ (light-green), and $k \in [4:31]$ for $1024^3$ (dark-green) resolution. The thick error bar indicates the uncertainty of the group scaling exponent and the thin error bar states the variation of the scaling exponent with time.
The results of different authors are shown with grey symbols, as stated in the legend. Additionally, the power law fits to the $256^3$ simulation (dotted), $512^3$ (dotted), and $1024^3$ (solid) are also shown with the fitting values stated in the legend.}
\label{fig:Resolution_comp}
\end{figure}

%%%%%%%%%%%%%%%%%%%%%%%%%%%%%%%%%
%% SUBSECTION 
%%%%%%%%%%%%%%%%%%%%%%%%%%%%%%%%%
\subsection{The influence of the Mach number and the resolution}
\label{subsec:MachInfluence}
In the next step we quantify the influence of the resolution. Therefore, we fit the power spectrum over an extended range starting at the largest scale, which is not directly influenced  by the forcing routine ($k=4$), and containing a number of points that doubles with the resolution. These ranges are $k \in [4:10]$, $k \in [4:17]$ and $k \in [4:31]$ containing $7$, $14$, $28$ data points for the $256^3$, $512^3$, and $1024^3$ simulation, respectively,
%This approach has several advantages.
%These fitting ranges, which increase with resolution, contain roughly the same scale dependent physical mechanisms influencing the density power spectrum for all resolutions. The only parameter, which we have to choose is \textit{one} width of the fitting range for the lowest or largest resolution. Another advantage is that we can directly compare our results with the works of other authors, as this approach results in 
which are comparable to those widely used \citep[e.g.][]{Kritsuk2007, Federrath2010, Konstandin2012a, Bertram2015}.
We follow the standard approach in the literature and fit a power law to the spectrum. As we see in Figure \ref{fig:Localslope_mix_dens}, the spectrum is actually slightly curved, and we note that this introduces additional uncertainty that we need to include in our Bayesian analysis. Therefore, we analyse the influence of increasing the uncertainty estimate artificially on our results as well as
varying the fitting range in the  Appendix \ref{sec:appendix}.

Figure \ref{fig:Resolution_comp} shows the scaling  exponents as a function of the Mach number for the simulations with $256^3$ (cyan), $512^3$ (light-green), and $1024^3$ (dark-green) resolution.
Additionally, we added the results of \citet{Kim2005} (grey circles), \citet{Kritsuk2007} (grey square) and \citet{Kowal2007a} (grey diamonds). The data-point of \citet{Kritsuk2007} ($1024^3$, compressive driven)
is in agreement with our results, whereas the data of \citet{Kim2005} ($512^3$, solenoidal driven) and \citet{Kowal2007a} ($256^3$, solenoidal driven, weak magnetic field) are systematically shallower for the highly supersonic cases. This is caused by the different forcing routines and confirms the finding that solenoidal forcing yields shallower density power spectra than mixed or compressive driven ones \citep{Federrath2009}. 
Another reason is the weak magnetic field in the simulations of \citet{Kowal2007a}, which is known to flatten the spectra further \citep{Padoan2004}.

To describe the functionality of the scaling exponent with the Mach number, we perform a Bayesian power law fit with two parameters $\zeta(\mathcal{M})=\alpha \mathcal{M}^{\beta}$ on the results with different resolutions.
We assume that $\alpha$, $\beta$, and the error on the measured scaling exponents are normally distributed.
The result of the regression is shown as solid ($1024^3$), dashed ($512^3$), and dotted ($256^3$) black lines and parameters are listed in the legend of Figure \ref{fig:Resolution_comp}.
With our value $\beta<0$ the model is in agreement with the theory of \citet{Saichev1996} as it converges towards the scaling $\mathcal{D} \propto k^0$ for the strong shock regime $\mathcal{M} \rightarrow \infty$. It also recovers the weak shock regime $\mathcal{D} \propto k^{-2}$ at the Mach number $\mathcal{M} = (-2/\alpha)^{1/\beta}$ and the scaling $\mathcal{D} \propto k^{-7/3}$ \citep{George1984, Bayly1992} at $\mathcal{M} = (-7/3\alpha)^{1/\beta}$.

To get an estimate of this functionality in the limit of infinite resolution, we 
assume a model with three parameters,
\begin{equation}
\zeta(\mathcal{M},\, n) = \alpha \mathcal{M}^{\tilde{\beta}+\omega \sum_{i=0}^n{\left(1/2\right)^i}}\, ,
\label{eq:model}
\end{equation}
where $n$ is a factor corresponding to the resolution $(2^n\, 256)^3$ of the simulation.
This model uses the assumption that the influence of the resolution on the measurement of the scaling exponent halves by doubling the resolution,
which is in agreement with our individual fits. We perform a Bayesian model fitting the three parameters of the scaling exponents of all resolutions simultaneously with the result
\begin{equation}
\alpha = -1.91 \pm 0.01,\, \tilde{\beta} = -0.70 \pm 0.01 ,\, \omega = 0.20 \pm 0.01\,.
\label{eq:model_res}
\end{equation}
This is in agreement with our individual fits shown in Figure  \ref{fig:Resolution_comp}.
The sum converges for $n\rightarrow \infty$ towards $2$ such that we get in the limit of infinite resolution $\beta = \tilde{\beta} +2 \omega $
\begin{equation}
\zeta(\mathcal{M}) = \left( -1.91 \pm 0.01 \right) \mathcal{M}^{-0.30\pm 0.03}\, .
\label{eq:model_conv}
\end{equation}
This is a remarkable result, as we can confirm for the first time that the trend of shallower slopes of the density power spectrum with increasing Mach number is independent of the resolution.

%###############################################################################
%Discussion
%###############################################################################
\section{Discussion}
\label{sec:discussion}
\subsection{The fractal dimension}
\label{sec:FractalDims}
In analogy to the hierarchical structure of the velocity, characteristic for incompressible turbulence theory, \citet{Weizsaecker1951} 
 introduced a hierarchy of clouds.
He proposed a theory describing the density distribution of molecular clouds
\begin{equation}
\frac{\rho_{\nu} }{\rho_{\nu-1} } =\left( \frac{\ell_{\nu} }{\ell_{\nu-1} } \right)^{-3\gamma} = f^{-1}\,,
\label{eq:Weizsaecker}
\end{equation}
where $\rho_{\nu}$ is the density of a cloud at the level of the hierarchy $\nu$, $\ell_{\nu}$ is the size of the cloud at this level,
$\gamma $ reflects the degree of compression, and $f$ is the volume filling factor. He assumes a self similar behaviour of the density field such that every cloud contains a certain number of smaller clouds and so on,
yielding density distributions described by equation (\ref{eq:Weizsaecker}). In this picture $\gamma$  is zero or one for no or isotropic compression, respectively.
\citet{Fleck1996} extended the work of \citet{Weizsaecker1951} and proposed a relation between the scaling of the density and the fractal dimension $D$, 
\begin{equation}
\rho(\ell) \propto \Theta(\ell) \ell^{-3 \gamma} \propto \Theta(\ell) \ell^{D-3}\,,
\label{eq:Fleck}
\end{equation}
where we added the unit step function $\Theta(\ell)$, as $\rho(\ell)$ is only defined for positive scales $\ell$. The Fourier transformation of equation (\ref{eq:Fleck}) gives
\begin{equation}
\rho(k) = \Gamma(1- 3\gamma ) k^{-\left( 1- 3 \gamma \right)}\,,
\label{eq:FT_Fleck}
\end{equation}
for the magnitude, neglecting the phase, and with $\Gamma$ being the Gamma function. Using the definition of the density spectrum equation (\ref{eq:def_spectrum}) and equation (\ref{eq:Theory}) we get
\begin{equation}
-\left( 1- 3\gamma \right)=\frac{1}{2}\zeta(\mathcal{M})\,,
\end{equation}
which leads to
\begin{equation}
\gamma =  \frac{1}{3}+\frac{1}{6} \alpha \mathcal{M}^{\beta}\,,
\label{eq:gamma}
\end{equation}
and finally results in
\begin{equation}
D =  2 - \frac{1}{2} \alpha \mathcal{M}^{\beta} \,.                                                                                                                                                                                                                                                                                                                                                                                                                                                                                                                                                                                                                                                                                                                                                                                                                                                                                                                                                        
\label{eq:D}
\end{equation}
Note that equation (\ref{eq:D}) is in agreement with the fractal dimension defined in \citet{Stutzki1998}, who
derived a relation between the scaling exponent of the density spectrum and the fractal drift exponent (Hurst exponent) based on the theory of fractal images. They get $D=E+1 -H$, with the dimensionality $E=2$ of the fractal surface and the Hurst exponent $H=\left(\zeta-2\right)/2$.
In the case of fractal images, $D$ can be interpreted as fractal box coverage dimension.
With equation (\ref{eq:D}) we get a fractal dimension of $D=2$ and $\gamma = 1/3$ in the limit $\mathcal{M} \rightarrow \infty$.
Note that this is a special case, as for $\gamma \rightarrow 1/3$ equation (\ref{eq:Fleck}) is $\Theta(\ell) \ell^{-1} = \delta(\ell)$ the Dirac Delta function, which Fourier transforms to constant magnitude $\propto k^0$ with a zero phase instead of equation (\ref{eq:FT_Fleck}), as $\Gamma(0)$ is not defined.

At low Mach numbers the average flow velocity is smaller than the sound speed such that it can not produce significant overdensities anymore.
To describe this transition, we define a critical Mach number $\mathcal{M}_{crit}$ below which we assume a constant density in equation (\ref{eq:Fleck}). The Fourier transformation in equations (\ref{eq:FT_Fleck}) and (\ref{eq:def_spectrum}) give a scaling exponent for the density spectrum of $\zeta(\mathcal{M}_{crit}) = -2$ in agreement with the theory of \citet{Saichev1996}.
Assuming $\gamma = 0$ and $D=3$  in equations (\ref{eq:gamma}) and  (\ref{eq:D}), we obtain
\begin{equation}
\mathcal{M}_{crit} = \left( \frac{-2}{\alpha}\right)^{1/\beta} = 0.86\,.
\end{equation}
using the measurements of equation (\ref{eq:model_conv}).
For Mach numbers $\mathcal{M} < \mathcal{M}_{crit} $ it follows that $\zeta<-2$ and $\gamma$ becomes negative. Therefore, $\gamma$ can not be interpreted as compression parameter anymore and from
equation (\ref{eq:Fleck}) we see that the density fluctuations are confined to the small scales. This regime is dominated by sound waves which have a  steep spectrum with $\zeta=-7/3$ \citep{George1984}.
The proposed range for the fractal dimension is also in agreement with observations \citep[e.g.][and references therein]{Elmegreen1996, Sanchez2007} and simulations \citep{Federrath2009, Konstandin2012a} suggesting an overall fractal dimension of interstellar clouds in the range $D \approx 2.0-2.7$.

%%%%%%%%%%%%%%%%%%%%%%%%%%%%%%
%% SUBSECTION
%%%%%%%%%%%%%%%%%%%%%%%%%%%%%%
\subsection{The curvature of the density power spectrum}
\label{subsec:Curved}
\begin{figure}
\includegraphics[width=0.98\linewidth]{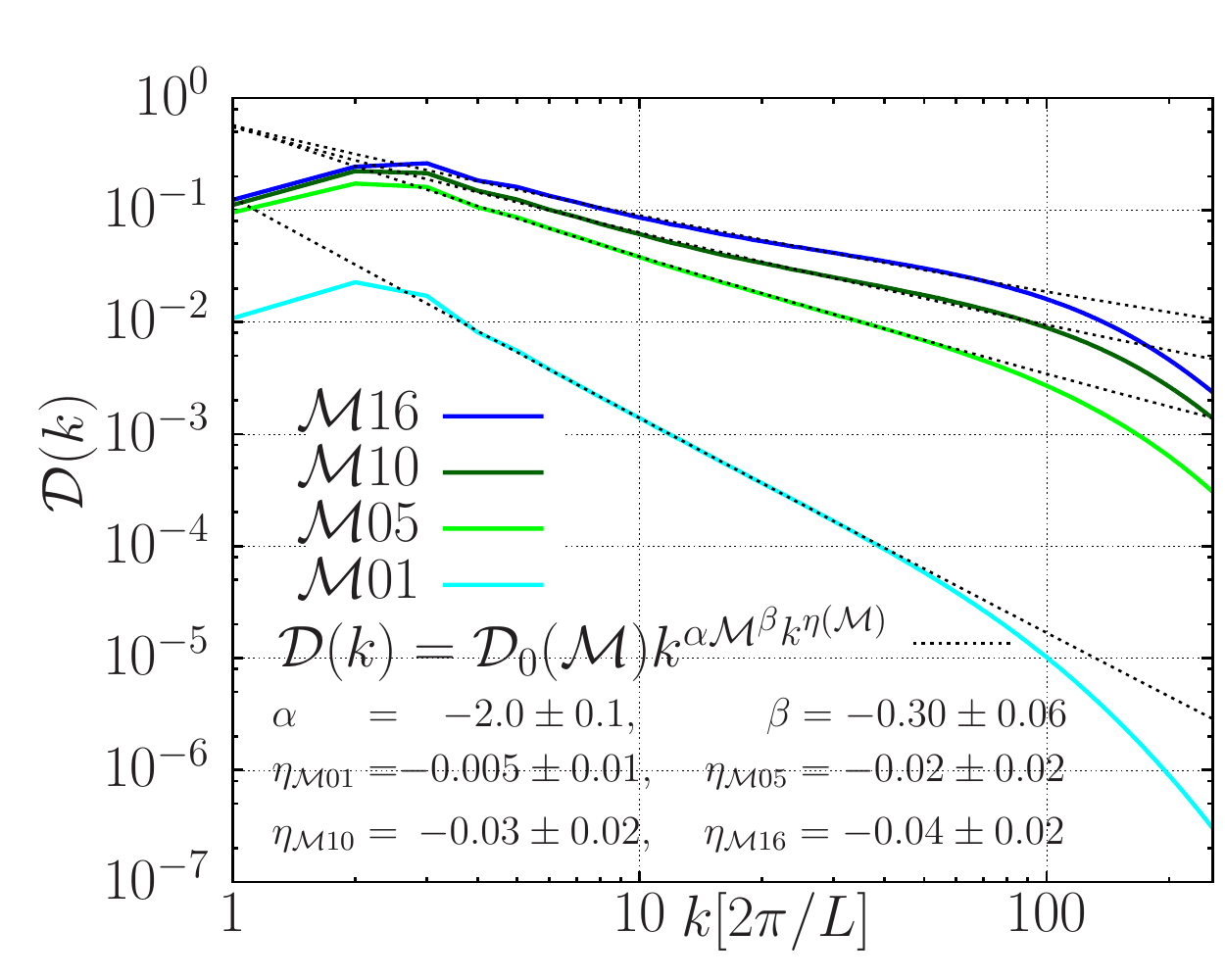}
\caption{Same as Figure \ref{fig:Spec_dens}, additionally showing the fit of equation (\ref{eq:CurvedPL}) as dotted lines. The fitting parameters of the scaling exponents are collected in the legend. Including the new curvature parameter $\eta $ to the model improved the fit significantly although $\eta $ is in most cases very small.}
\label{fig:new_model}
\end{figure}
The second conclusion concerns the density spectrum itself.
The discussion in section \ref{sec:results} has shown that the scaling exponents of the density spectrum are not only changing in between the simulations with different Mach numbers (Figure \ref{fig:Resolution_comp}) but also
in each simulation with the scale (Figure \ref{fig:Localslope_mix_dens}).
Considering a driving force producing a turbulent flow with the Mach number $\mathcal{M}_0=V_0/c_s$ at the scale $\ell_0$ and
assuming a power law scaling $\ell^{p}$ of the velocity fluctuations, there is a scale $\ell_s$ at which the velocity fluctuations are of the order of the sound speed $c_s$. This scale is called the sonic scale \citep{Vazquez2003}. According to our analysis in Section 
\ref{subsec:MachInfluence} we expect a scaling exponent of the density spectrum of $\zeta=\alpha \mathcal{M}_0^{\beta}$ above the sonic scale.
Increasing the Mach number of the box to $\mathcal{M}' > \mathcal{M}_0$ yields $V'/c_s = \left( \ell_0/ \ell_s' \right)^p > \left( \ell_0/ \ell_s \right)^p = V_0/c_s$ after reaching a statistically steady state, resulting in a smaller sonic scale $\ell_s > \ell_s'$ and a shallower density spectrum
$\zeta  < \zeta'$. However, we can find a scale $\tilde{\ell}'$ given by  $( \tilde{\ell}'/ \ell'_s )^p \overset{!}{=} \left( \ell_0/ \ell_s \right)^p =\mathcal{M}_0 $.
This implies that the turbulent flow with higher Mach number $\mathcal{M}'$ contains a sub volume of size $\tilde{\ell}'$ with the Mach number 
$\mathcal{M}_0 $, the same dynamical range relative to the sonic scale, and therefore the same scaling exponent $\zeta$ as the original box, while $\zeta' = \alpha (\mathcal{M'})^{\beta} > \zeta$ for the new box of size $\ell_0$.
This example illustrates that we have to include a dependence on the scale in our model of the scaling exponent, which we will do in the following paragraph.

Guided by this picture and the variation of the local scaling exponents with the scale, we make a power law ansatz for the scaling exponent
 $\zeta(\mathcal{M},\,k) =\zeta(\mathcal{M}) k^{\eta}$. We neglect this variation with the scale in Section \ref{sec:results}, which can be justified as we investigate in equations (\ref{eq:model}) and (\ref{eq:model_conv}) the scaling behaviour of the simulations only in a limited range close to the forcing regime.
Using the scaling exponent $\zeta(\mathcal{M})=\alpha \mathcal{M}^{\beta}$ established in Section \ref{subsec:MachInfluence} and add the scaling factor $k^{\eta}$ to account for a curved spectrum results in
\begin{equation}
\mathcal{D}(k) = \mathcal{D}_0(\mathcal{M}) k^{\alpha \mathcal{M}^{\beta} \,k^{\eta(\mathcal{M})}}\,.
\label{eq:CurvedPL}
\end{equation}
The fit  in $k\in [4:31]$ of the density spectra with $1024^3$ is shown in Figure (\ref{fig:new_model}) and the resulting fitting parameter for the scaling exponents are collected in its legend. 
We performed a hierarchical Bayesian model fitting all spectra with different Mach numbers at once such that $\alpha$ and $\beta$ do not depend on the Mach number and are consistent with the measurements in Section \ref{subsec:MachInfluence}.
The additional factor $k^{\eta}$ improved the fit significantly such that the fit and the density spectra are hardly distinguishable for $k<80$. The fitting parameter $\eta $ describes the deviation from a pure power law, which is nearly zero for the low Mach number simulation, such that in this case the density power spectrum is well described by a single power law. For the high Mach number cases the parameter $\eta $ increases up to $ -0.04 \pm 0.02$. We interpret the measurements of $\eta$ as lower limits of the curvature, as
the density spectra get still steeper at small wave numbers $k\approx 7$ with resolution, whereas their scaling exponents at intermediate scales changes only slightly with resolution (Figure \ref{fig:Localslope_mix_dens}, or Figure \ref{fig:Mach_vs_Slope_approaches} panel in the middle and on the right).

\begin{figure}
\includegraphics[width=0.98\linewidth]{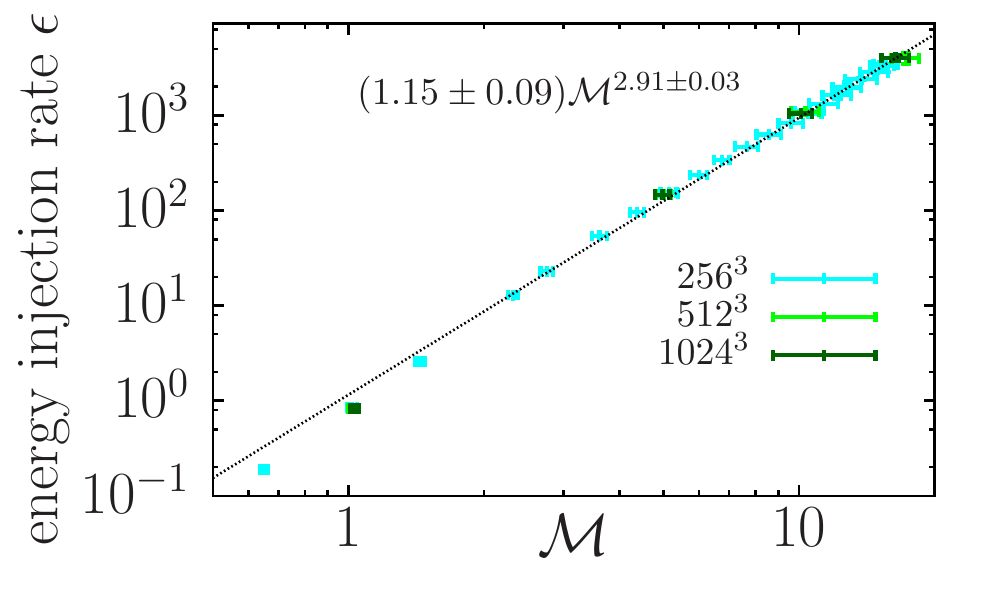}
\caption{The energy injection rate $\epsilon$ as a function of the Mach number $\mathcal{M}$ with a power law fit. The resulting fitting parameters are given in the figure.}
\label{fig:Mach_vs_DEinj}
\end{figure}

Before we close this section, we want to mention that the parametrisation of our simulations with the Mach number is arbitrary and we chose it as it is used in the derivation of \citet{Saichev1996} and most studies of supersonic turbulence \citep[e.g.][]{Kim2005, Schmidt2006, Konstandin2012a}. We can change the parametrisation to any quantity that has a bijective mapping to the Mach number. Figure \ref{fig:Mach_vs_DEinj} shows as examples the energy injection rate of our forcing routine, which we fitted with $\epsilon \propto \mathcal{M}^{2.91}$.
Other possibilities are the standard deviation of the density, which is $\sigma_{s}^2= \mathrm{b} \log{(1+ \mathrm{b}^2\mathcal{M}^2)}$ (shown in Figure \ref{fig:Mach_vs_rho}) or the compressive Mach number as discussed in \citet{Konstandin2012b}.
Expressing the scaling exponent of the density spectrum with the energy injection rate (via its influence on the Mach number, see Figure \ref{fig:Mach_vs_DEinj}) has two advantages. First,
it measures the total energy flowing through the cascade as the turbulent boxes are in statistical steady state and shows explicitly that we do not need the Mach number as additional parameter
to describe a supersonic turbulent flow. Hence, our results for the density spectrum of supersonic turbulence suggest that KomogorovÕs second hypothesis holds, which states that all small scale statistical properties are uniquely and universally determined by the scale $\ell$ and the mean energy dissipation rate $\epsilon$ \citep{Kolmogorov1941a,Kolmogorov1941b}. The second advantage is that it offers additional interpretations of our results.

%###############################################################################
% Summary
%###############################################################################
\section{Summary}
\label{sec:summary}
We analyse the properties of turbulence using a suite of three-dimensional numerical simulations which are continuously driven on the largest scales. The forcing scheme consists both solenoidal (transverse) and compressive (longitudinal) modes in equal parts. We model driven, compressible, isothermal, turbulence with rms Mach numbers ranging from the subsonic to the highly supersonic regime.
We find the relation $\sigma_{s}^2 = \mathrm{b}\log{(1+\mathrm{b}^2\mathcal{M}^2)}$ between the Mach number and the standard deviation of the density distribution, which improves the fit significantly. We derive this relation with the shock jump condition and the Fokker-Planck equation (Section \ref{sec:DensityPDF}). We find $\mathrm{b}= 0.457 \pm 0.007$ with the new proposed formula describing the mixture of compressive and solenoidal modes of the velocity field, which is in agreement with our driving scheme.
By employing a hierarchical Bayesian fitting method, we estimate the parameters describing the scaling relation of the density power spectrum.
The density power spectra follow power laws, $\mathcal{D} \propto k^{\zeta(\mathcal{M})}$, with a scaling exponent
depending on the Mach number (Section \ref{subsec:MachInfluence}) in agreement with the theory of \citet{Saichev1996}.
We find that $\zeta(\mathcal{M})= \alpha \mathcal{M}^{\beta}$ with $\alpha$ scattering slightly with resolution, whereas $\beta$ gets systematically shallower. 
We model that effect and extrapolate to the limit of infinite resolution (equation \ref{eq:model}) to find $\zeta(\mathcal{M}) = \left( -1.91 \pm 0.01 \right) \mathcal{M}^{-0.30\pm 0.03}$. 
We validate this result by testing the influence of varying position and width of the fitting range, as well as the uncertainty of measured scaling exponents of the density spectrum on the inferred parameters (Appendix \ref{sec:appendix}). 

The dependence of the scaling exponent on the average Mach number of the density spectrum implies a dependence of the fractal dimension on the Mach number (Section \ref{sec:FractalDims}). In the proposed model the fractal dimension is $D=2-1/2 \,\zeta(\mathcal{M})= 2 + 0.96 \mathcal{M}^{-0.30}$. The fractal dimension is $D=2$ in the strong shock regime and $D=3$ in the incompressible limit, which is reached at the critical Mach number $\mathcal{M}_{crit} \approx 0.86$. This is in agreement with the observations of \citet[][and references therein]{Elmegreen1996} suggesting an overall fractal dimension of interstellar clouds in the range $ D \approx 2.0-2.7$. 

We also determine how the parameters depend on the wavenumber and quantify the deviation from a pure power law by moving the fitting range systematically over the density spectrum (Section \ref{sec:localSE}). This analysis reveals that the density power spectra are slightly curved. This curvature gets more pronounced with increasing Mach number. The density spectra are steeper close to the forcing scale, shallow at intermediate scales and again steeper on small scales. The height of this bump in the local scaling exponents increases with the Mach number.

We develop a physically motivated fitting formula reproducing the deviations from a pure power law based on the Mach number dependence of the scaling exponent of the density power spectrum 
 (Section \ref{subsec:Curved}). 
We propose $\mathcal{D}(k) = \mathcal{D}_0 k^{\zeta k^{\eta}}$ with $\zeta = \alpha \mathcal{M}^{\beta}$ and a new parameter ${\eta}$ describing the deviation of the spectrum from a pure power law with fixed scaling exponent. This functionality describes all density spectra down to wave numbers of $k \approx 80$.   
We measure ${\eta} = -0.005 \pm 0.01$ in the low Mach number regime such that in this case the density power spectrum follows a single power-law, and we find the strongest curvature ${\eta} = -0.04 \pm 0.02$ in the simulation with the highest Mach number.

%%%%%%%%%%%%%%%%%%%%%%%
%%% APPENDIX          %
%%%%%%%%%%%%%%%%%%%%%%%
\appendix
\section{Stability of the fitting results}
\label{sec:appendix}
\begin{table*}
\caption{Results of the three parameter fit ($\alpha,\, \tilde{\beta},\, \omega$) for different estimates of the uncertainty of the local scaling exponents (first four columns) and for different fitting ranges (fifth and sixth column).}
\label{tab:fittingresults}
\begin{tabular}{@{} l c c c c c c  }
                           \multicolumn{5}{c}{$k \in [4:10]_{256}$, $[4:17]_{512}$, $[4:31]_{1024}$ } & $k \in [4:10]_{\mathrm{all}}$ &$k \in [4:10]_{256}$, $[11:17]_{512}$,\, $[25:31]_{1024}$\\
                           &   $\sigma_{\zeta}$   &  $\sigma_{t}$         & $20\% \,\zeta$       & $0.15$                   & $\sigma_{\zeta}$  & $\sigma_{\zeta}$ \\
\hline
$\alpha$           &   $-1.91  \pm 0.01 $ & $-1.85  \pm 0.01$  & $-1.93 \pm 0.14$   & $-1.90  \pm 0.06$ &$-1.87  \pm 0.01$ & $-1.88 \pm 0.01$ \\
$\tilde{\beta}$  &   $-0.70  \pm 0.01 $ & $-0.65  \pm 0.03$  & $-0.71 \pm 0.06$   & $-0.69  \pm 0.06$ &$-0.75  \pm 0.01$ & $-0.60 \pm 0.01$ \\
$\omega$        &   $ 0.20  \pm 0.01 $ & $ 0.19  \pm 0.02$  & $  0.21 \pm 0.05$   & $ 0.20  \pm 0.05$ & $ 0.26 \pm 0.01$ & $ 0.11 \pm 0.01$\\
\hline
$\beta$              &   $-0.30  \pm 0.03$ & $-0.27 \pm  0.07$  & $-0.29 \pm  0.16$   & $-0.29 \pm  0.16$ &$-0.23 \pm 0.03$ & $-0.38 \pm 0.03$\\
\hline
\end{tabular}
\end{table*}

\begin{figure*}
\includegraphics[width=0.98\linewidth]{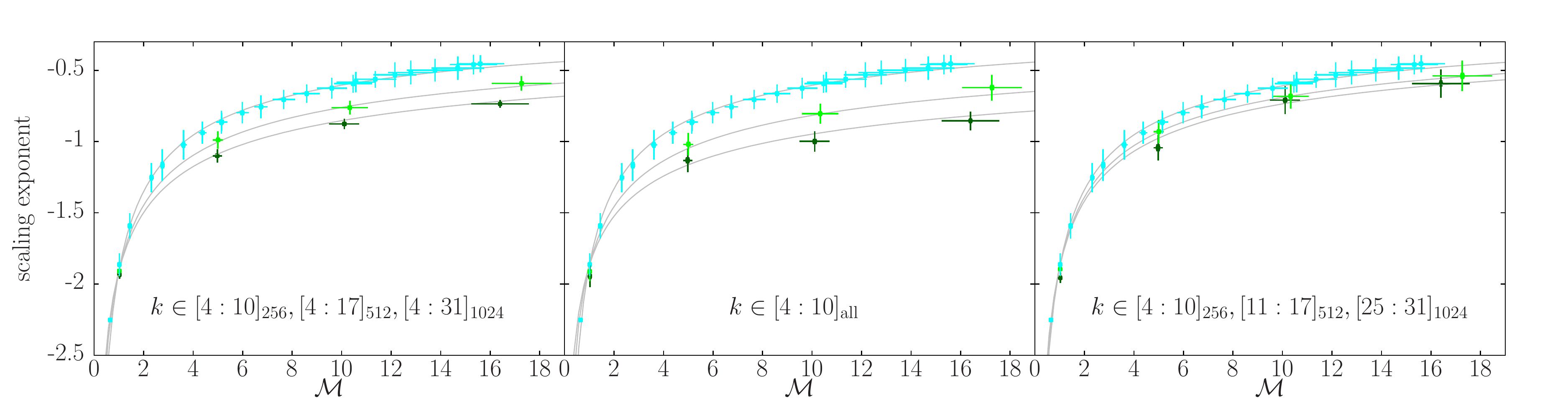}
\caption{Scaling exponents as a function of the Mach number $\mathcal{M}$ for the $256^3$ (cyan), $512^3$ (light-green), and $1024^3$ (dark-green) resolution simulations. The scaling exponents are measured in varying fitting ranges as indicated in the figure. Additionally, the result of three-parameter fit of equation (\ref{eq:model}) collected in Table \ref{tab:fittingresults} (columns $2,\,5,\,6$) are shown as grey lines from left to right.}
\label{fig:Mach_vs_Slope_approaches}
\end{figure*}

As stated in the main part, the approach of choosing an extended fitting range, which doubles with the resolution fits a power law function, where the density spectrum is slightly curved.
This could lead to wrong uncertainty estimates of $\zeta(\mathcal{M})$, as the assumption of a power law is only approximately true especially for the higher Mach numbers. 
Therefore, we test in the following how stable the presented result is against increasing the uncertainty of $\zeta(\mathcal{M})$.
The results are collected in Table (\ref{tab:fittingresults}).
The first row contains the result already presented in equation (\ref{eq:model_res}).
Using the standard deviation of the time variation $\sigma_{t}$ as the uncertainty results in parameter estimates shown in the second, estimating it with $20\%$ of the measured scaling exponent is in the third, and with a constant $0.15$ is in the fourth column.
The last two columns contain results, where we analyse the influence of the fitting range. Therefore, we use a constant width of the fitting range $\Delta k = 6 $ for all resolutions and the error estimate $\sigma_{\zeta}$.
The fifth column contains the result, in which the scaling exponents are measured with a fitting range $[4:10]$ independent of the resolution, whereas the last column contains the result fitting the flat, shallow parts in Figure \ref{fig:Localslope_mix_dens},
so $[4:10],\, [11:17],\,[25:31]$ centred at $7$, $14$, $28$ for the $256^3$, $512^3$, and $1024^3$ resolution, respectively.
The parameter estimates in Table (\ref{tab:fittingresults}) show that the presented results are independent of small variations.
They also confirm the finding that the scaling exponent of the density power spectrum follows $\zeta(\mathcal{M}) \propto \mathcal{M}^{\approx -0.30}$ in the limit of infinit resolution.
Figure \ref{fig:Mach_vs_Slope_approaches} shows the scaling exponents and the fitting curves of the parameters in the  columns $2,\,5,\,6$
of Table \ref{tab:fittingresults} to demonstrate the goodness of the fits.

%%%%%%%%%%%%%%%%%%%%%%%
%%% ACKNOWLEDGMENTS   %
%%%%%%%%%%%%%%%%%%%%%%%
\section*{Acknowledgments}
L.K.~acknowledge financial support by the European Research Council for the FP7 ERC starting grant project LOCALSTAR.
L.K.~and R.S.K.~gives thanks to the Deutsche Forschungsgemeinschaft (DFG) for financial support via the SFB 881 'The Milky Way System' (subproject B1, B2, B5) as well as the SPP 1573 'Physics of the Interstellar Medium',
and via the individual grant KL1358/11.
R.S.K.~furthermore acknowledges funding from the European Research Council under the European Community's Seventh Framework Programme (FP7/2007-2013) via the ERC Advanced Grant ``STARLIGHT: Formation of the First Stars'' (project number 339177).
P.G.~acknowledges the support by the Max-Planck Institut f\"{u}r Astrophysik in Garching, support from the DFG Priority Program 1573 \textit{Physics of the Interstellar Medium} and acknowledges the support of a Marie Curie Research Training Network (MRTN-CT2006-035890).
T.P.acknowledges financial support from the DFG priority program 1537 ÔPhysics of the interstellar mediumÕ.
We acknowledge computing time at the Leibniz-Rechenzentrum (LRZ) in Garching under project ID h1343.
The software used in this work was in part developed by the DOE-supported ASC / Alliance Center for Astrophysical
Thermonuclear Flashes at the University of Chicago. The Bayesian models for the parameter estimates were produced using BAT \citep{Caldwell2009}.\\

%%%%%%%%%%%%%%%%%%%%%%%
%%% BIBLIOGRAPHY      %
%%%%%%%%%%%%%%%%%%%%%%%
\bibliography{thesis}
\bibliographystyle{mn2e}

\label{lastpage}
\end{document}